\documentclass[]{mosi}

\usepackage{helvet}
\usepackage{array}
\usepackage{colortbl}
\usepackage{float}
\usepackage{longtable}
\usepackage{tabularx}
\usepackage{needspace}
\usepackage{listings}
\usepackage{textcomp}
\usepackage{xspace}
\usepackage{algorithm}
\usepackage{algpseudocode}
\usepackage{multicol}
\usetikzlibrary{arrows.meta,positioning,fit}

\usepackage{amsmath,amsfonts,bm}

\def\eqref#1{equation~\ref{#1}}

\def\1{\bm{1}}

\DeclareMathAlphabet{\mathsfit}{\encodingdefault}{\sfdefault}{m}{sl}
\SetMathAlphabet{\mathsfit}{bold}{\encodingdefault}{\sfdefault}{bx}{n}

\newcommand{\apptablestyle}{%
  \normalsize
  \renewcommand{\arraystretch}{1.16}%
  \setlength{\tabcolsep}{4pt}%
}

\title{VehicleArena: A Realistic Urban Environment\\for Multi-Agent Driving}
\hypersetup{
  pdftitle={VehicleArena: A Realistic Urban Environment for Multi-Agent Driving},
  pdfauthor={Jie Yang, Jiajun Chen, Jiazheng Zhou, Mianqiu Huang, Yining Zheng, Yuxin Wang, Xipeng Qiu}
}

\author[1,*]{Jie Yang}
\author[1,2,*]{Jiajun Chen}
\author[1,*]{Jiazheng Zhou}
\author[1]{Mianqiu Huang}
\author[1,2]{Yining Zheng}
\author[2,\dagger]{Yuxin Wang}
\author[1,2,\dagger]{Xipeng Qiu}

\affiliation[1]{Fudan University}
\affiliation[2]{Shanghai Innovation Institute}

\abstract{Real-world embodied agents often pursue independent objectives within a shared physical environment, where their actions can alter the conditions faced by others. Existing benchmarks, however, typically assume shared goals or explicitly prescribed interaction protocols, leaving such emergent physical coupling underexplored. We introduce \textbf{VehicleArena}, a 3D urban-driving benchmark for studying independently operating agents in a dynamic shared world. In VehicleArena, LLM-controlled agents must fulfill evolving passenger requests while navigating complex traffic, and each agent's driving decisions can reshape traffic flow, delays, risks, and subsequent observations for surrounding agents. The benchmark provides 112 evaluation tasks---80 single-agent and 32 multi-agent. Across nine evaluated models, the highest arrival rates reach only 65.0\% on single-agent tasks and 65.6\% on multi-agent tasks, while strong passenger-request or cabin scores do not reliably translate into successful trip completion. Moreover, in matched multi-agent runs, every evaluated driving policy reduces the arrival rate of surrounding vehicles relative to the simulator's native traffic controller, revealing measurable externalities beyond the ego vehicle itself.
}
\checkdata[Github]{\url{https://github.com/OpenMOSS/VehicleArena}}
\checkdata[Dataset]{\url{https://huggingface.co/datasets/OpenMOSS-Team/VehicleArena}}

\begin{document}
\maketitle
\begingroup
\renewcommand{\thefootnote}{\fnsymbol{footnote}}
\setcounter{footnote}{1}
\footnotetext{Equal contribution.}
\setcounter{footnote}{2}
\footnotetext{Corresponding authors: Yuxin Wang (\href{mailto:wangyuxin@sii.edu.cn}{wangyuxin@sii.edu.cn}) and Xipeng Qiu (\href{mailto:xpqiu@sii.edu.cn}{xpqiu@sii.edu.cn}).}
\endgroup

\begin{figure*}[!t]
\centering
\includegraphics[width=\textwidth]{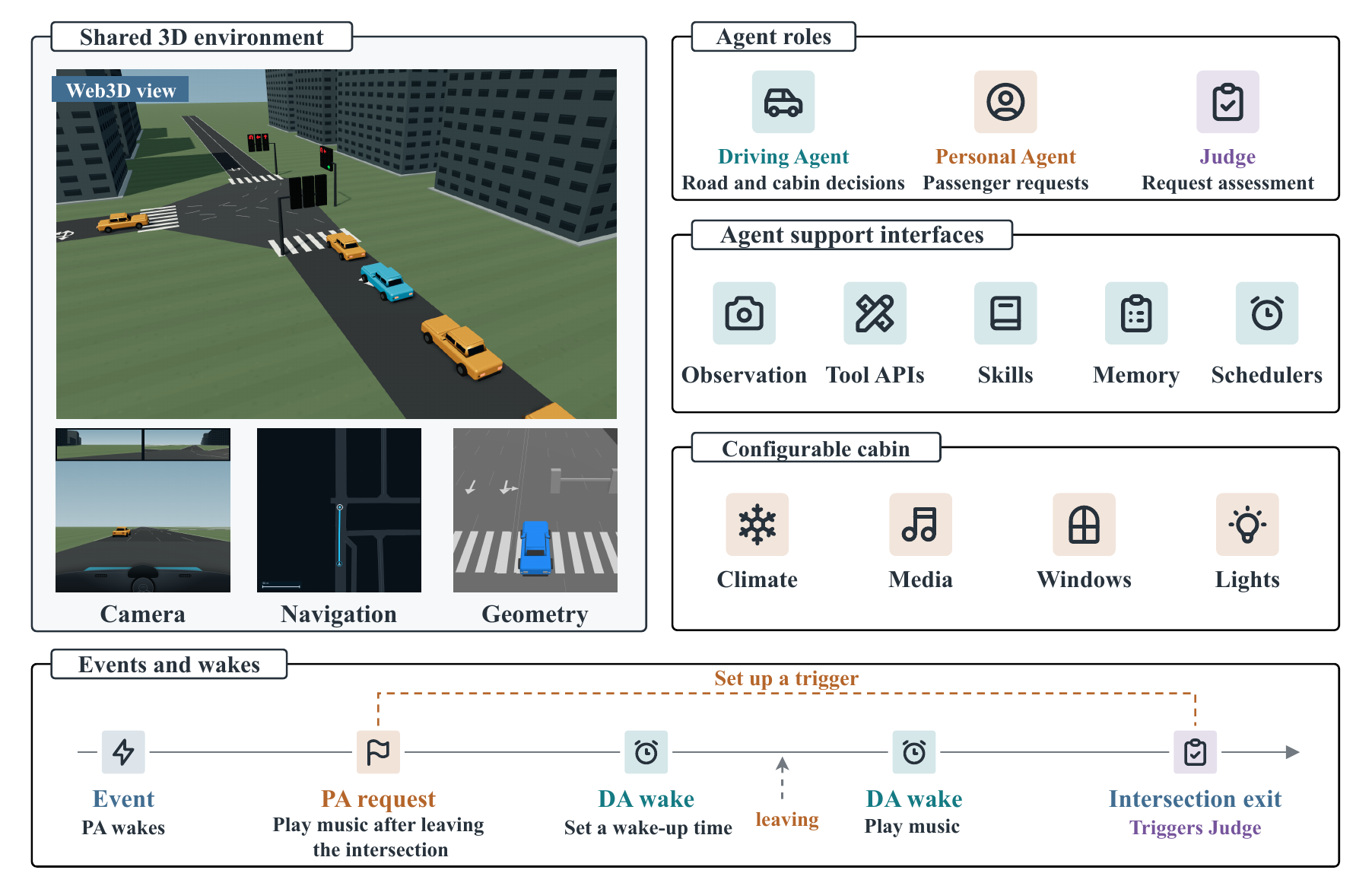}
\caption{VehicleArena architecture. Left: the shared 3D world and observation examples. Right: agent roles, support interfaces, and configurable cabin capabilities. Bottom: PA asks DA to play music after leaving an intersection and registers a verification trigger; DA sets a wake-up time to carry out the request, while Judge checks are triggered independently of DA wakes. The timeline is illustrative and not to scale. The main image is a native Web3D capture from a SUMO-only reference rollout; the smaller images are separate observation examples.}
\label{fig:overview}
\end{figure*}

\section{Introduction}
\label{sec:introduction}

Language and vision-language agents are increasingly being studied in embodied driving settings, yet existing work has largely progressed along two separate tracks. Closed-loop driving benchmarks focus on on-road decision making, evaluating an ego vehicle's multimodal perception, planning, and instruction-following capabilities~\citep{shao2024lmdrive,yang2025drivearena,fu2024limsim++}. In parallel, intelligent cockpit environments emphasize in-cabin interaction and device control~\citep{yang2025vehicleworld}. Although both settings capture important agent capabilities, they are typically evaluated in isolation: passenger intent and cabin interaction are separated from the physical driving process. This makes it difficult to study how an agent should respond to evolving passenger requests while continuously adapting to a changing road environment.

To bridge this gap, we introduce VehicleArena, a high-fidelity 3D urban-driving benchmark that unifies passenger interaction, cockpit control, and physical driving within a single closed loop. VehicleArena integrates passenger-oriented Personal Agents with LLM-controlled Driving Agents. Personal Agents issue requests involving routes, driving preferences, and in-vehicle devices, while Driving Agents must interpret and execute these requests under dynamic traffic conditions. As a result, passenger intent, cabin operations, route decisions, and vehicle motion evolve together along a shared timeline.

This unified design naturally extends to multi-vehicle settings. VehicleArena supports multiple LLM-controlled vehicles operating concurrently in the same urban environment, with each vehicle serving its own passenger and pursuing its own objectives. Unlike conventional multi-agent benchmarks, where agent relationships are typically defined through explicit collaboration, competition, or communication structures~\citep{li2023camel,zhou2024sotopia,zhu2025multiagentbench,wang2024battleagentbench}, agents in VehicleArena need not share goals, plans, or interaction protocols. Instead, they influence one another through the physical consequences of their actions in a shared road environment.

A route change, merge, yield, sudden stop, or hesitation by one vehicle can immediately alter the observations, risks, and feasible actions of surrounding agents. We refer to this setting as independent-objective physical coexistence: independently motivated agents become coupled not through a predefined social structure, but through persistent physical interaction in a shared world. This setting allows us to ask a central question: can LLM agents remain reliable and safe when their behavior must account for both changing user demands and the unpredictable consequences of other independently acting agents?

To support research in this setting, VehicleArena provides 100 development tasks and a held-out evaluation set of 112 tasks. The evaluation set includes 80 single-agent tasks that assess integrated driving and passenger-request execution, and 32 MultiLLM interaction tasks that probe multi-agent physical coexistence. We evaluate Driving Agents using metrics spanning trip completion, driving quality, passenger-request satisfaction, cabin correctness, traffic externalities, and inference cost.

Our contributions are summarized as follows:
\begin{itemize}
\item We develop a high-fidelity 3D agentic-driving environment that unifies multimodal driving, passenger-side objectives, intelligent cockpit interaction, and dynamic urban traffic within a single closed loop.

\item We extend the environment to multi-vehicle scenarios and formulate independent-objective physical coexistence, where agents with independent objectives interact through the physical consequences of their actions in a shared world.

\item We construct a benchmark with 112 held-out evaluation tasks covering both integrated single-agent capabilities and multi-agent physical coexistence, together with metrics for driving, passenger-request execution, traffic impact, and inference cost.
\end{itemize}

\section{Related Work}
\label{sec:related}

\subsection{Multi-Agent LLM Systems}

Multi-agent LLM systems often organize agents around a shared objective. CAMEL uses role-playing dialogue, while ChatDev and MetaGPT assign specialized roles and workflows for collaborative software development~\citep{li2023camel,qian2024chatdev,hong2024metagpt}. PARTNR extends this paradigm to embodied human--robot collaboration~\citep{chang2025partnr}. Other benchmarks introduce private or competitive objectives: SOTOPIA studies agents with individual social goals, while MultiAgentBench and BattleAgentBench cover cooperation and competition under predefined interaction structures~\citep{zhou2024sotopia,zhu2025multiagentbench,wang2024battleagentbench}. Across these settings, agent relationships are largely specified by the task, whether as teammates, opponents, or negotiation partners.

VehicleArena instead studies \emph{physical coexistence among agents with independent objectives}. Agents serve different users and need not share goals, plans, or communication protocols. Their interactions emerge through a shared physical environment: a lane change, yield, or sudden stop by one vehicle changes the risks and feasible actions of nearby agents. VehicleArena therefore focuses on whether independently operating agents remain reliable when their coupling arises from persistent physical interaction rather than a predefined social structure.

\subsection{LLM-Based Autonomous Driving}

Recent language- and vision-language-based driving methods primarily study scene understanding and ego-vehicle decision making. SGDrive structures driving knowledge for VLM-based planning~\citep{li2026sgdrive}, while NAVSIM and Pseudo-Simulation enable scalable evaluation through non-reactive rollouts~\citep{dauner2024navsim,cao2025pseudo}. DriveBench further evaluates the visual grounding and robustness of VLM-based driving reasoning~\citep{xie2025drivebench}. These settings expose important perception and planning failures, but do not execute decisions in a persistent, reactive traffic world.

Closed-loop simulators such as CARLA, SMARTS, MetaDrive, and SUMO provide interactive traffic dynamics~\citep{dosovitskiy2017carla,zhou2020smarts,li2022metadrive,lopez2018microscopic}, and systems including LMDrive, DriveArena, and LimSim++ integrate multimodal LLMs into such environments~\citep{shao2024lmdrive,yang2025drivearena,fu2024limsim++}. However, they remain largely centered on an ego vehicle following a predefined navigation objective or instruction. In parallel, VehicleWorld studies LLM interaction with in-vehicle devices, but separates cockpit interaction from physical driving~\citep{yang2025vehicleworld}.

VehicleArena unifies these settings in a persistent environment where passenger requests, cabin state, driving decisions, and surrounding traffic evolve together. It further supports multiple independently controlled vehicles, connecting integrated passenger--driver interaction with the physical multi-agent coexistence described above.
\section{VehicleArena Environment}
\label{sec:method}

VehicleArena brings passenger service and physical driving into a shared urban world. We first introduce the system components, and then describe the harnesses that connect agent decisions, physical execution, and request verification.

\begin{figure}[!t]
\centering
\includegraphics[width=\textwidth]{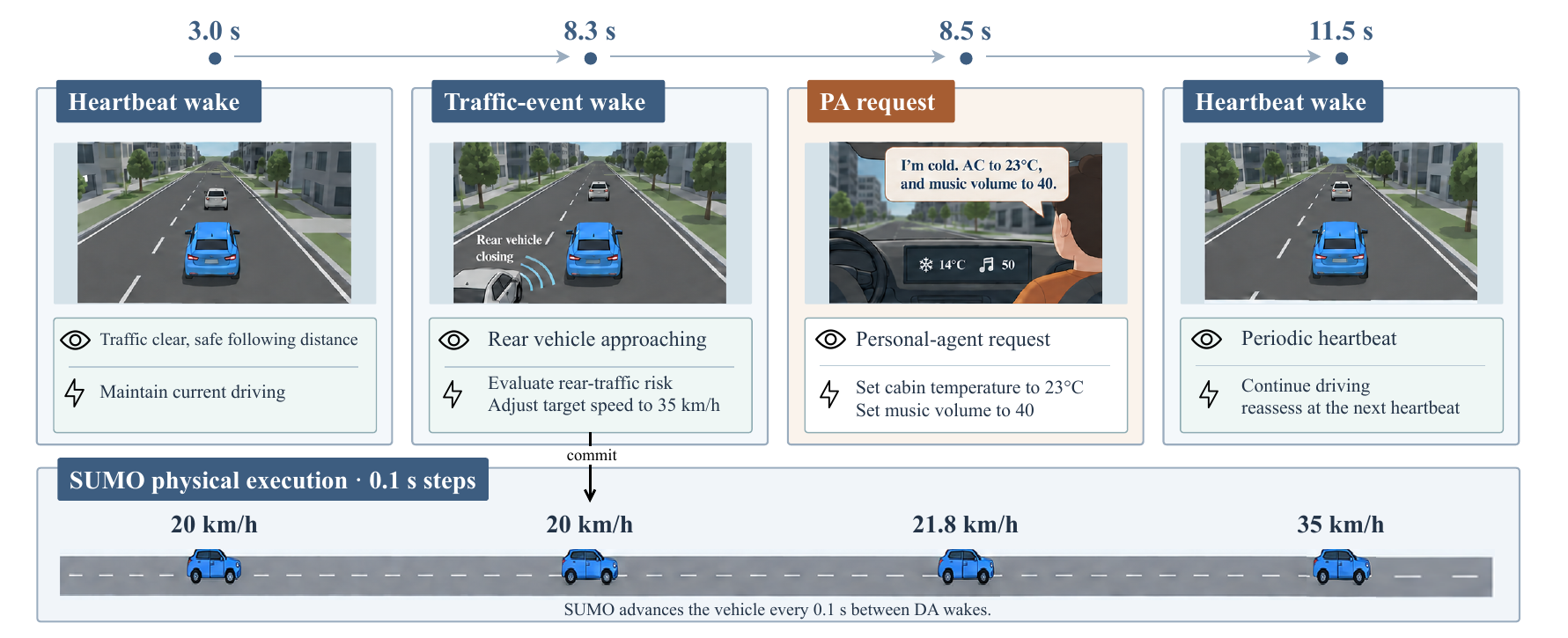}
\caption{A recorded episode illustrating interleaved driving and passenger service. The upper panels show three DA wakes and a passenger request, together with the relevant observations and actions. Following a rear-traffic warning, DA raises its target speed to 35\,km/h; it later handles a cabin request without replacing that motion target. The lower road shows the corresponding speeds realized by SUMO, which advances at 0.1-second steps between wakes. Time gaps are compressed and intervening wakes are omitted.}
\label{fig:driving-agent-loop}
\end{figure}

\subsection{Overview}
\label{sec:integrated-driving}
\label{sec:configuration-rules}
\label{sec:shared-environment}
\label{sec:npc-method}

Figure~\ref{fig:overview} illustrates a shared 3D environment in which traffic, cabin state, and environmental events evolve along a single simulation timeline. SUMO advances vehicle motion and traffic interactions, while synchronized rendering exposes the resulting world state to the agents. Vehicles can differ in physical capabilities and installed equipment; extensible cabin modules and a rule engine define available device operations and operating constraints. Each participating vehicle can be controlled either by an LLM-based driver or by native SUMO logic, allowing selected background vehicles to become independent agents without changing the surrounding environment. Vehicle configuration is detailed in Appendix~\ref{app:configuration-rules}.

Three agent roles operate over the same evolving episode. A Personal Agent (PA) expresses passenger needs, a Driving Agent (DA) determines how to serve them while driving, and a Judge evaluates whether execution satisfies the request. All three refer to the same world state and simulation clock, but have distinct responsibilities and information access: PA cannot directly actuate the vehicle, while Judge observes and evaluates execution without controlling it. In multi-vehicle scenarios, each DA serves its own passenger and destination. Their objectives remain independent, while their actions become physically interdependent through the shared road environment.
Appendix~\ref{app:prompts} presents their prompt templates; Appendix~\ref{app:tools} lists the tools available to each role.

\subsection{Driving-Agent Harness}
\label{sec:driving-harness}

\textbf{Agent loop.}
DA interacts with the world through repeated observe--act sessions, as illustrated in Figure~\ref{fig:driving-agent-loop}. At each wake, the harness combines current observations and passenger updates with retained task context. DA may inspect the situation, invoke tools, and use their feedback over multiple turns before ending the session. Recent interaction history and a persistent Todo board preserve unfinished work across wakes.

DA is not invoked at every physics step. In addition to wakes triggered by startup or relevant events, it controls a periodic heartbeat and may schedule a future wake, making \emph{when to reason} part of its policy. Between wakes, the world continues to evolve under the currently active controls. When multiple DAs wake at the same simulation time, they observe the same committed road state and run in parallel; their buffered motion commands are committed before the next physics step.

\textbf{Tool discovery and skills.}
The harness exposes capabilities progressively so that the same agent loop can operate across differently equipped vehicles. Core driving tools are immediately available, while additional capabilities can be discovered by inspecting installed modules and loading their corresponding interfaces. Optional \emph{skills} provide procedural guidance for multi-step operations without prescribing a driving strategy. DA can therefore discover and invoke the capabilities required by a passenger request during an ongoing episode.
The skill catalog and complete guide contents are provided in Appendix~\ref{app:skills}.

\textbf{Driving action space.}
DA controls the vehicle through persistent target-speed commands, ordinary or emergency braking, adjacent-lane changes, and junction maneuvers such as turning or continuing straight. Route planning provides navigation guidance rather than automatic route following; DA remains responsible for selecting the required maneuvers and signaling. These commands specify intended control rather than instantaneous state changes. The simulator realizes them under the vehicle's physical constraints and surrounding traffic, so braking takes time and a requested lane change may be constrained by nearby road users. Cabin tools similarly modify persistent device state rather than merely producing textual responses.

\subsection{Personal-Agent and Judge Harness}
\label{sec:pa-judge-method}

PA wakes on passenger-visible events or randomized timers. It observes outstanding requests and currently feasible trigger options, and may issue, revise, or cancel both immediate and delayed requests. Once a request is accepted, its desired outcomes and acceptance criteria are fixed. DA receives the passenger update but independently determines how to execute the request and when to wake again.

For delayed requests, the simulator monitors the trigger selected by PA and schedules Judge checks independently of DA wakes. Judge evaluates the fixed criteria against recorded interactions and physical execution history, incorporating earlier checks when a requirement must hold over time. A request remains pending until completion or its final verification point, while cases that cannot be reliably validated remain unscored. The complete loop is shown in Algorithm~\ref{alg:pa-judge-loop} in Appendix~\ref{app:pa-scheduling}; scoring details are provided in Section~\ref{sec:benchmark-metrics}.

\section{Benchmark}
\label{sec:benchmark}

The benchmark uses the system in Section~\ref{sec:method} to evaluate whether agents can complete their own trips and passenger tasks while sharing the road with others. We describe how tasks are constructed, how matched SUMO runs provide a comparison and time budget, and which outcomes are reported.

\subsection{Task Suite and Construction}
\label{sec:task-suite}
\label{sec:task-construction}

The suite contains 100 training and development tasks and 112 held-out evaluation tasks. The evaluation set comprises 80 \textsc{Basic} tasks with one DA-controlled vehicle and 32 \textsc{MultiLLM} tasks with several independently configured DAs. Appendices~\ref{app:task-suite} and~\ref{app:cities} summarize the task split and supported cities.

Tasks are constructed under human expert supervision, with experts participating in the design of traffic situations, vehicle trips, and passenger goals. The resulting tasks cover intersections, merges, narrow roads, and pedestrian crossings, requiring agents to complete their trips and respond to passenger requests.

\subsection{Evaluation Protocol}
\label{sec:evaluation-protocol}

\paragraph{Oracle baseline and time budget.}
We treat SUMO's native driving policy as a driving oracle. Before evaluating a DA, we run each scene with SUMO controlling the ego vehicle to establish reference driving behavior, traffic outcomes, and a task deadline. The baseline and evaluated runs use exactly the same environment and surrounding-agent configurations; only the ego vehicle's controller changes. In \textsc{MultiLLM}, the other LLM-controlled vehicles use Qwen3.8-27B with distinct personality prompts and active PAs, all held fixed across the two runs. This oracle baseline provides a reference for assessing both the DA's own driving behavior and its effects on other road users, including additional collisions, non-arrivals, and delays. To set the deadline, we record the last arrival time $T_{\mathrm{cal}}$ among required SUMO-controlled vehicles, excluding persistent obstacles and without waiting for LLM-controlled peers to arrive. With $T_{\mathrm{event}}$ denoting the last scheduled environmental event, we set
\begin{equation}
T_{\mathrm{limit}}=\max\left(T_{\mathrm{cal}},\;T_{\mathrm{event}}\right)+\Delta.
\label{eq:reference-limit}
\end{equation}
The margin $\Delta$ is normally 10 seconds. The resulting deadline is fixed throughout evaluation.

\subsection{Evaluation Metrics}
\label{sec:benchmark-metrics}

We report six complementary outcomes: trip completion and driving quality for the ego vehicle, passenger-request fulfillment and cabin compliance for service, effects on background traffic relative to the SUMO baseline, and model-token use for efficiency. These outcomes remain separate rather than being combined into one score. Table~\ref{tab:benchmark-metrics} in Appendix~\ref{app:metric-definitions} gives their definitions and reporting units; Appendices~\ref{app:metrics} and~\ref{app:driving-rules} detail passenger grading and driving-rule deductions.

\section{Experiments}
\label{sec:experiments}
\subsection{Experimental Setup}

We evaluate nine DA models on 112 held-out tasks: 80 \textsc{Basic} and 32 \textsc{MultiLLM} scenarios. Each model is evaluated three times per task. In \textsc{Basic}, only the ego vehicle uses the evaluated DA; other traffic participants retain native simulation control. In \textsc{MultiLLM}, the ego DA shares the road with three other DA-controlled vehicles. Each peer uses Qwen3.8-27B with a distinct personality prompt and an active PA. Their model, prompt, and PA assignments remain the same across model comparisons, although their actions respond to the evolving traffic.

The main comparison uses the same task IDs and scoring rules for each model. Traffic-impact analyses pair evaluated runs with SUMO references by task and required vehicle IDs.
Model specifications are listed in Appendix~\ref{app:model-signatures}.

\subsection{Main Results}

Table~\ref{tab:main-results} compares ego-vehicle outcomes in the two task groups; effects on surrounding vehicles are analyzed separately below.

\begin{table}[!htbp]
\caption{Main evaluation by task group. Basic has 80 tasks per model and Multi-Agent has 32. The best value in each metric column is bold; the second-best value is underlined, including ties.}
\label{tab:main-results}
\centering
\normalsize
\setlength{\tabcolsep}{1.5pt}
\begin{tabular*}{\textwidth}{@{\extracolsep{\fill}}lcccccccc@{\hspace{0.12in}}}
\toprule
 & \multicolumn{4}{c}{Basic (80 tasks)} & \multicolumn{4}{c}{Multi-Agent (32 tasks)} \\
\cmidrule(lr){2-5}\cmidrule(l){6-9}
Model & Arr. (\%) & Drive & Req. & Cabin & Arr. (\%) & Drive & Req. & Cabin \\
\midrule
DeepSeek-V4.1-Flash & \underline{62.5} & \textbf{85.0} & \underline{88.5} & \underline{76.6} & 56.3 & 81.7 & 82.1 & \underline{73.9} \\
MIMO-V2.6-Pro & 40.0 & 79.5 & \textbf{90.0} & 57.5 & \underline{62.5} & 80.0 & \underline{88.0} & 66.7 \\
Kimi-K3 & 51.3 & 78.9 & 84.9 & 41.5 & 40.6 & 71.4 & 76.4 & 49.2 \\
GLM-5.3-Flash & 45.0 & 78.6 & 85.7 & 55.6 & \underline{62.5} & 80.9 & 85.7 & 63.1 \\
Qwen3.8-Max & \textbf{65.0} & \underline{83.1} & 86.7 & 72.0 & \textbf{65.6} & \underline{82.4} & \textbf{88.2} & 65.4 \\
Qwen3.8-27B & 36.3 & 79.6 & 83.1 & 47.4 & 40.6 & 74.0 & 86.7 & 54.5 \\
GPT-5.6-Sol & 20.0 & 82.5 & 80.8 & \textbf{90.2} & 9.4 & \textbf{87.2} & 78.1 & \textbf{94.0} \\
Qwen3-VL-8B & 17.5 & 39.7 & 39.0 & 11.6 & 18.8 & 44.5 & 40.0 & 7.0 \\
Qwen3-VL-32B & 21.3 & 49.3 & 60.2 & 34.3 & 25.0 & 53.3 & 58.2 & 31.8 \\
\bottomrule
\end{tabular*}
\end{table}

\textbf{Basic shows a wide spread in ego-vehicle arrival.} Qwen3.8-Max and DeepSeek-V4.1-Flash reach 65.0\% and 62.5\% arrival, respectively, while MIMO-V2.6-Pro has the highest Request score at 90.0. GPT-5.6-Sol has the highest Cabin score at 90.2 but reaches only 20.0\% arrival. This pattern is consistent with insufficient progress before the task deadline despite strong cabin performance.

\textbf{Multi-Agent results show a similar separation between arrival and other scores.} Qwen3.8-Max reaches 65.6\% arrival, followed by MIMO-V2.6-Pro and GLM-5.3-Flash at 62.5\% each. GPT-5.6-Sol records the highest Drive and Cabin scores, 87.2 and 94.0, but reaches only 9.4\% arrival. Across all tasks it also has the lowest mean ego speed (11.4\,km/h) and no recorded collisions (Table~\ref{tab:model-physical-performance}), consistent with limited progress before the task deadline.

Across both task groups, the order of models changes across the four metrics: strong driving, passenger-request, or cabin scores do not consistently coincide with successful arrival. Reporting these outcomes separately exposes the gap between completing a trip and satisfying other parts of the task.

\subsection{Analysis}

We inspect the ego vehicle's recorded decisions to characterize how each model spends inference and tool calls, and how it drives. We report Basic and Multi-Agent tasks separately for the quantitative comparisons. For the plots, we combine the four metrics in Table~\ref{tab:main-results}: ego arrival, Drive, Request, and Cabin. We average the latter three scores, then weight that average by the ego vehicle's arrival rate. This completion-weighted index captures trip completion and task quality on one outcome axis for comparing planning effort; Appendix~\ref{app:completion-weighted-index} gives the exact formula.

\begin{figure}[!htbp]
\centering
\includegraphics[width=0.90\textwidth]{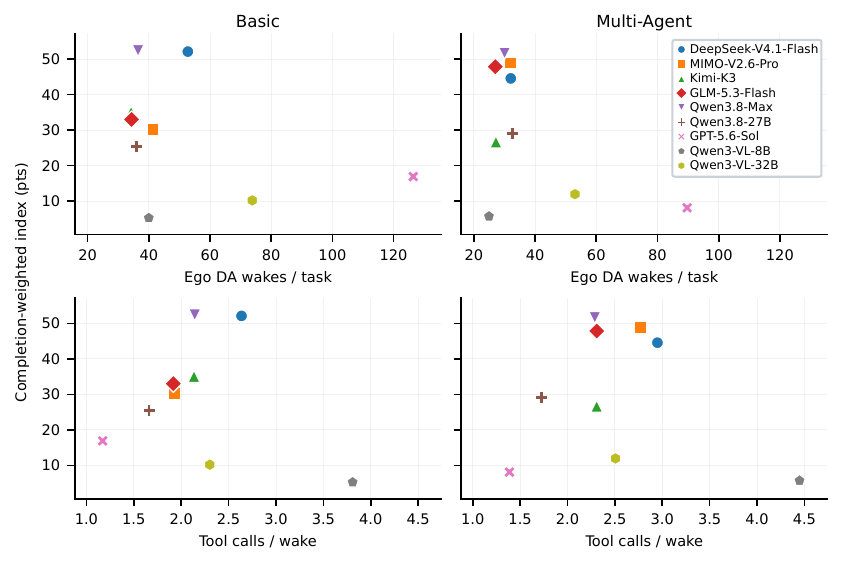}
\caption{Planning activity versus completion-weighted outcome for Basic (left) and Multi-Agent (right): ego DA wakes per task (top) and tool calls per wake (bottom). Each marker denotes a model; Appendix~\ref{app:completion-weighted-index} defines the outcome index.}
\label{fig:model-planning-efficiency}
\end{figure}

\textbf{Planning efficiency varies across models.} We compare models within each task group by outcome and interaction cost. A short failed run is not good planning merely because it uses little interaction. Figure~\ref{fig:model-planning-efficiency} shows wake frequency and tool calls per wake; together these determine tool calls per task. Qwen3.8-Max illustrates the favorable pattern: on Basic it reaches nearly the same outcome as DeepSeek-V4.1-Flash with fewer wakes and fewer tool calls per task, and it remains strong on Multi-Agent. GLM-5.3-Flash and MIMO-V2.6-Pro have similar Multi-Agent outcomes, but GLM uses fewer wakes and fewer calls within each wake. GLM-5.3-Flash and Kimi-K3 have similar wake frequencies, tool calls per wake, and outcomes on Basic. Their interaction counts remain similar on Multi-Agent, but GLM achieves a much stronger outcome than Kimi. This contrast shows that interaction volume alone does not explain their different outcomes in shared traffic.

GPT-5.6-Sol illustrates how planning activity can become unproductive: repeated wakes with few tool calls per wake accumulate without strong outcomes. Wake count also depends on episode duration and events, and call count says nothing about whether an observation or action was timely.

Overall, Qwen3.8-Max pairs strong outcomes with economical interaction, while frequent shallow wakes do not ensure success. GLM and Kimi show that similar interaction counts can yield different outcomes. Appendix~\ref{app:supplementary-analyses} extends this analysis with skill, tool, and token statistics; Section~\ref{app:token-cost-analysis} compares token use with the same outcome index.

\textbf{Speed and collisions characterize physical driving.} Table~\ref{tab:model-physical-performance} pools all tasks for each model. Qwen3.8-Max has the highest mean speed, few ego collisions, and the strongest arrival in both task groups. MIMO-V2.6-Pro and GPT-5.6-Sol have no ego collisions, but MIMO moves faster and arrives more often. GLM-5.3-Flash and Kimi-K3 have similar mean speeds, yet GLM reaches substantially more Multi-Agent destinations. Their Basic arrival rates are closer, which the pooled speed cannot reveal. These contrasts show that neither speed nor collision count alone captures effective driving.

\begin{table}[H]
\centering
\caption{Ego vehicle speed and collisions across all 112 tasks. Mean speed includes stopped time; collisions count ego-involved events.}
\label{tab:model-physical-performance}
\normalsize
\setlength{\tabcolsep}{2pt}
\renewcommand{\arraystretch}{1.1}
\begin{tabular*}{0.7\textwidth}{@{\extracolsep{\fill}}lcc@{}}
\toprule
Model & Speed (km/h) & Collisions \\
\midrule
DeepSeek-V4.1-Flash & \underline{19.0} & 4 \\
MIMO-V2.6-Pro & 16.8 & 0 \\
Kimi-K3 & 18.6 & 4 \\
GLM-5.3-Flash & 18.5 & 3 \\
Qwen3.8-Max & \textbf{19.2} & 3 \\
Qwen3.8-27B & 16.9 & 5 \\
GPT-5.6-Sol & 11.4 & 0 \\
Qwen3-VL-8B & 14.3 & 27 \\
Qwen3-VL-32B & 14.8 & 18 \\
\bottomrule
\end{tabular*}
\end{table}

\begin{figure}[!htbp]
\centering
\includegraphics[width=\textwidth]{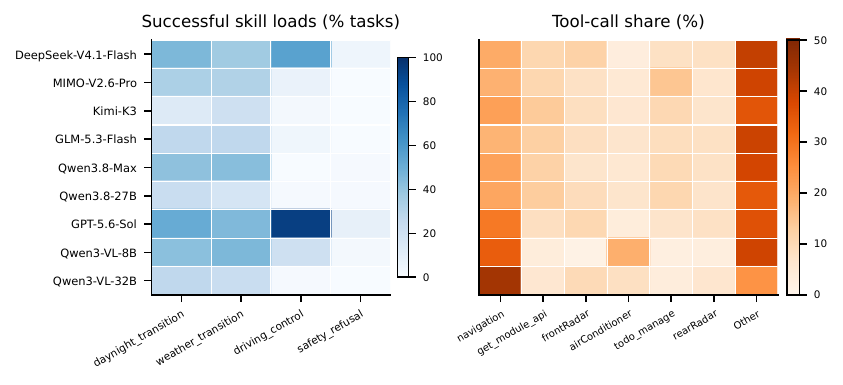}
\caption{Ego DA skill and tool choices pooled over Basic and Multi-Agent tasks. Both heatmaps share the model order shown on the left. Left: fraction of tasks with a successful load of each skill. Right: share of non-\texttt{finish} tool calls in the six most common tool families; ``Other'' contains the remainder. Each panel has its own percentage color scale.}
\label{fig:model-skill-tool}
\end{figure}

\textbf{Skill and tool choices differ across models.} In Figure~\ref{fig:model-skill-tool}, blue cells show the percentage of 112 tasks with a successful skill load; orange cells show each family's percentage of non-\texttt{finish} tool calls. GPT-5.6-Sol has the largest summed skill-load frequency, including \emph{driving\_control} in 105 tasks, 96 at the initial destination wake, matching the guide's navigation trigger. It also makes many \texttt{navigation} calls, mainly minimap and target-speed requests, yet has low arrival and completion-weighted outcomes. Qwen3.8-Max navigates effectively without loading that guide, so loading it is not required for effective tool use. Appendix~\ref{app:model-preference-counts} gives the counts.

\subsection{Impact on Surrounding Vehicles}

\begin{figure}[!t]
\centering
\includegraphics[width=\textwidth]{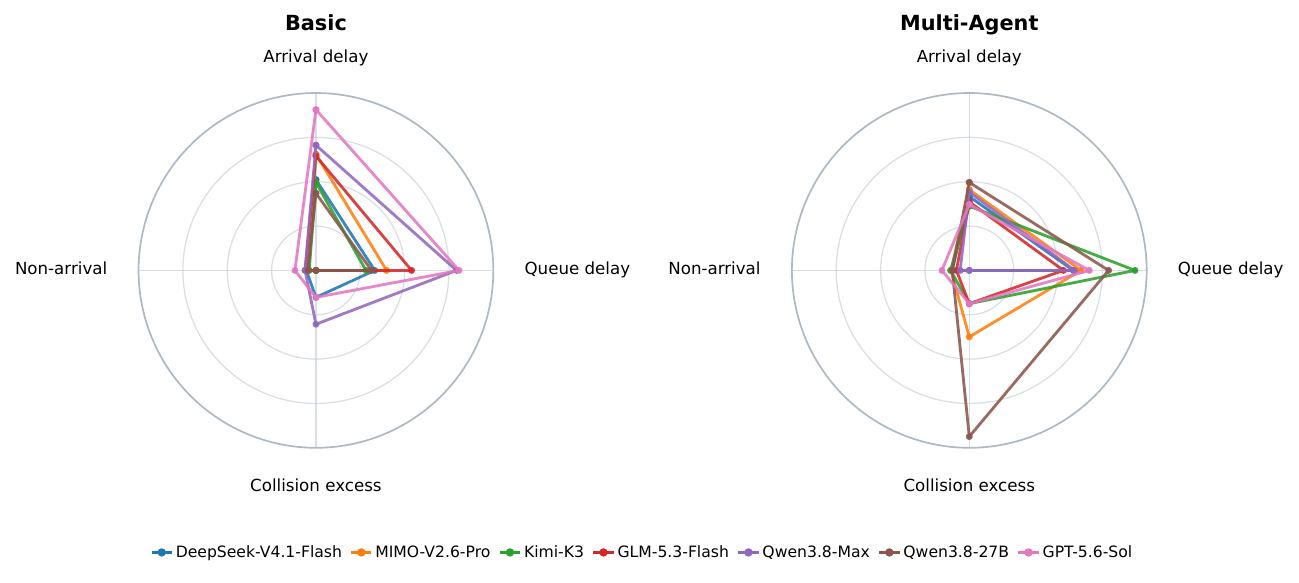}
\caption{Surrounding-traffic burden relative to matched SUMO runs in Basic (left) and Multi-Agent (right). Axes show added delay and collision/non-arrival rate excess; reductions appear as zero. Appendix~\ref{app:surrounding-impact} gives definitions and scales.}
\label{fig:surrounding-impact}
\end{figure}

\textbf{The ego vehicle's behavior also shapes surrounding traffic.} Figure~\ref{fig:surrounding-impact} compares required non-ego trip vehicles, including fixed peers, with same-task SUMO runs. It shows added delay and changes in collision and non-arrival rates; Appendix~\ref{app:surrounding-impact} defines the metrics.

\textbf{Surrounding traffic bears different costs across models and task groups.} Across the seven models shown, the equal-weight mean of the four axes across both groups is largest for GPT-5.6-Sol and smallest for DeepSeek-V4.1-Flash. On Basic, GPT has the largest burden, chiefly from delays and missed trips, consistent with its slow, often unfinished ego trips in Table~\ref{tab:model-physical-performance}; Qwen3.8-27B has the smallest. On Multi-Agent, Qwen3.8-27B has the largest burden, including the highest collision-rate increase and substantial queue delay. Qwen3.8-Max has the smallest burden and the fewest added non-arrivals despite visible delays. Although its surrounding-vehicle collision rate is lower than SUMO's, we plot this favorable change as zero collision excess because the radar emphasizes the adverse effects that predominate across models.

\begin{table}[!t]
\caption{Ego PA ablation using the main-table aggregation. Each scored cell reports the PA-off value and its change from PA on in parentheses; arrival changes are percentage points.}
\label{tab:pa-ablation}
\centering
\normalsize
\setlength{\tabcolsep}{1.5pt}
\begin{tabular*}{\textwidth}{@{\extracolsep{\fill}}lcccccc@{\hspace{0.12in}}}
\toprule
 & \multicolumn{3}{c}{Basic (80 tasks)} & \multicolumn{3}{c}{Multi-Agent (32 tasks)} \\
\cmidrule(lr){2-4}\cmidrule(l){5-7}
Model & Arr. (\%) & Drive & Cabin & Arr. (\%) & Drive & Cabin \\
\midrule
Kimi-K3 & 61.3 ($+10.0$) & 74.6 ($-4.3$) & 71.6 ($+30.1$) & 56.3 ($+15.6$) & 77.2 ($+5.8$) & 81.6 ($+32.4$) \\
GLM-5.3-Flash & 62.5 ($+17.5$) & 79.1 ($+0.5$) & 75.6 ($+20.0$) & 65.6 ($+3.1$) & 78.8 ($-2.1$) & 83.3 ($+20.2$) \\
\bottomrule
\end{tabular*}
\end{table}

\subsection{Ablation}

We disable the ego vehicle's Personal Agent and Passenger Judge while retaining both components for the fixed peer agents. Table~\ref{tab:pa-ablation} reports the resulting changes.
Disabling ego PA raises overall arrival by 11.6 percentage points for Kimi-K3 (48.2\% to 59.8\%) and 13.4 points for GLM-5.3-Flash (50.0\% to 63.4\%). Across both task groups, field-weighted Cabin scores increase by 30.9 and 20.2 points, respectively.

\textbf{In-cabin requests compete with time-critical tasks.} With PA enabled, ego acts on passenger device requests while missing simultaneous mandatory cabin checks in 23/112 Kimi-K3 tasks (20.5\%) and 21/112 GLM tasks (18.8\%). This exposes difficulty coordinating concurrent obligations.

\textbf{Passenger preferences versus trip completion.} Among tasks completed only with ego PA disabled, 9/21 for Kimi-K3 (42.9\%) and 10/26 for GLM (38.5\%) show a low-speed passenger request, matching speed command, and PA-on timeout without collision or route failure. Table~\ref{tab:pa-low-speed-examples} in Appendix~\ref{app:pa-low-speed-examples} gives examples, suggesting passenger preferences can delay arrival.

\section{Conclusion}
\label{sec:conclusion}

VehicleArena studies independent agents whose private plans meet in a shared physical world with persistent, observable consequences. Its 3D traffic world, passenger requests, cabin actions, and non-compensatory metrics connect local decisions to system-level effects. Across nine models, strong passenger-request or cabin scores often coexist with failed trips, while the evaluated driving policies can reduce other vehicles' arrival rates. Future work should test longer episodes, richer sensing, broader road cultures, and transfer to real driving data.

\subsection*{Acknowledgments}

We gratefully acknowledge Shanghai Qiji Zhifeng Co., Ltd. for its support of this project, which helped make this research possible.

\bibliographystyle{plainnat}
\bibliography{references}

\appendix
\setlength{\LTpre}{6pt}
\setlength{\LTpost}{8pt}
\setlength{\LTcapwidth}{\columnwidth}

\section{Experimental Settings}
\label{app:experimental-settings}

\subsection{Inference Configuration}

\paragraph{Role-specific inference.}
Table~\ref{tab:inference-settings} summarizes the requested inference settings
and documented thinking/effort controls. Effort labels are model-specific;
N/A marks controls not exposed by the model. Model specifications are listed
in Appendix~\ref{app:model-signatures}.

\begin{table}[!htb]
\caption{Recorded inference settings for the main comparison. Context budgets
are in tokens; 32k denotes 32,768 output tokens per call.}
\label{tab:inference-settings}
\centering
\normalsize
\setlength{\tabcolsep}{3pt}
\begin{tabular*}{\columnwidth}{@{\extracolsep{\fill}}llcccrr@{}}
\toprule
Role & Model & Temperature & Thinking & Effort & \shortstack{Context\\budget} & \shortstack{Output\\limit} \\
\midrule
\multirow{9}{*}{Ego DA} & DeepSeek-V4.1-Flash & 0.7 & Enabled & max & 1000000 & 32k \\
& MiMo-v2.6-Pro & 1 & Enabled & N/A\textsuperscript{a} & 1000000 & 32k \\
& Kimi-K3 & 0.7 & Enabled & max & 1000000 & 32k \\
& GLM-5.3-Flash & 0.7 & Enabled & max & 1000000 & 32k \\
& Qwen3.8-Max & 0.7 & Enabled & xhigh & 1000000 & 32k \\
& Qwen3.8-27B & 0.7 & Enabled & xhigh & 1000000 & 32k \\
& GPT-5.6-sol & 0.7 & Enabled & high & 1000000 & 32k \\
& Qwen3-VL-8B & 0.7 & N/A\textsuperscript{b} & N/A & 1000000 & 32k \\
& Qwen3-VL-32B & 0.7 & N/A\textsuperscript{b} & N/A & 1000000 & 32k \\
Peer DA & Qwen3.8-27B & 0.7 & Enabled & xhigh & 1000000 & 32k \\
PA & Qwen3.8-27B & 0.7 & Enabled & xhigh & 1000000 & 32k \\
Judge & Qwen3.8-27B & 0.0 & Enabled & xhigh & 1000000 & 32k \\
\bottomrule
\end{tabular*}
\par\vspace{3pt}
\begin{minipage}{\columnwidth}
\normalsize
\textsuperscript{a}MiMo does not expose adjustable effort.
\textsuperscript{b}Qwen3-VL uses instruct checkpoints.
\end{minipage}
\end{table}


In \textsc{MultiLLM}, peer vehicles retain their distinct personality prompts
and active PAs and Judges in both the oracle baseline and the evaluated run. Only the
ego vehicle changes controller; all other experimental settings are matched.
The ego PA/Judge ablation disables those two ego roles while retaining
the fixed peers' configuration. Without the ego PA, no ego passenger requests
are generated. Cabin scoring uses the same automatic rule path in both settings
(Section~\ref{app:cabin-rule-evaluation}).

\subsection{Task Split}
\label{app:task-suite}

\paragraph{Split and counting conventions.}
Table~\ref{tab:task-suite-statistics} summarizes 100 \textsc{Basic} training
tasks and 112 test tasks: 80 \textsc{Basic} and 32 \textsc{MultiLLM} tasks
selected by the frozen run manifests. The Basic training pool contains the
100 scenarios outside the Basic test manifest. The table counts all LLM-controlled vehicles,
including peers in each \textsc{MultiLLM} task. Map counts are deduplicated
within each split, so aggregate map counts are not sums of the component splits.

\begin{table}[!htb]
\caption{Task-suite statistics. Tasks and maps are counts; other entries are
per-task means, with ranges in parentheses.}
\label{tab:task-suite-statistics}
\centering
\normalsize
\setlength{\tabcolsep}{2pt}
\renewcommand{\arraystretch}{1.18}
\begin{tabular*}{\columnwidth}{@{\extracolsep{\fill}}lcccc@{}}
\toprule
Statistic & Train & \shortstack{Test\\\textsc{Basic}} & \shortstack{Test\\\textsc{MultiLLM}} & Overall \\
\midrule
Tasks & 100 & 80 & 32 & 212 \\
Maps & 57 & 48 & 13 & 89 \\
\addlinespace[2pt]
LLM vehicles & 1.0 & 1.0 & 4.0 & 1.5 \\
NPC vehicles & 13.1 (8--24) & 13.2 (8--24) & 10.3 (8--13) & 12.7 (8--24) \\
Pedestrians & 0.5 (0--8) & 0.5 (0--8) & 0.0 & 0.4 (0--8) \\
Ego route (m) & 373.8 (22.0--2620.7) & 402.4 (63.4--4867.2) & 199.6 (75.8--427.9) & 358.3 (22.0--4867.2) \\
Time limit (s) & 88.0 (20.3--514.3) & 89.8 (21.7--490.6) & 50.0 (24.6--73.9) & 82.9 (20.3--514.3) \\
\bottomrule
\end{tabular*}
\par\vspace{3pt}
\begin{minipage}{\columnwidth}
\normalsize
Ego route is the lane-level distance from the initial pose to the destination,
using the frozen route or the lane planner when no initial route is stored.
Time limit is the calibrated simulation deadline in Equation~\ref{eq:reference-limit}.
\end{minipage}
\end{table}

\paragraph{Task families.}
\textsc{Basic} covers crosswalk yielding, signalized and unsignalized
intersections, straight following, required and continuous turns, full-network
navigation, lead-vehicle braking, narrow-road interaction, red-light stopping,
oncoming and crossing streams, unprotected left turns, obstacle-driven gap
changes, platoon pressure, and merge streams. \textsc{MultiLLM} combines
narrow meetings, unprotected left turns, synchronized four-way intersections,
multi-vehicle merges, and dense four-way and merge variations. Across both
splits, weather can be stable or transition among cloudy, foggy, rainy, snowy,
heavy-rain, heavy-snow, and hail conditions; illumination includes dawn,
morning, noon, afternoon, dusk, and night. Passenger requests are generated
online by PA under the configuration described in Section~\ref{app:pa-scheduling}.

\subsection{PA and Judge Scheduling}
\label{app:pa-scheduling}

\subsubsection{Personal-Agent wakes}

The PA input includes passenger-visible state, recent motion, the wake
category, prior requests, and passenger-visible DA replies. Judge grades,
hidden evaluation state, and DA private reasoning are excluded.

\begin{table}[!htb]
\caption{Default PA scheduling. The random stream is independently seeded
per vehicle and scenario.}
\label{tab:pa-wakes}
\centering
\normalsize
\begin{tabular}{@{}>{\raggedright\arraybackslash}p{45mm}@{\hspace{3mm}}>{\raggedright\arraybackslash}p{\dimexpr\columnwidth-48mm\relax}@{}}
\toprule
Wake source & Default rule \\
\midrule
Shared passenger-visible events & Simulation start, acoustic cue, weather change, and day--night change. \\
Hard braking edge & Acceleration at most $-3\,\mathrm{m/s^2}$ for 0.3 seconds; detection is rearmed only after acceleration exceeds $-1.5\,\mathrm{m/s^2}$. \\
Prolonged-stop edge & Speed below $0.5\,\mathrm{km/h}$ for 15 seconds; detection is rearmed after speed exceeds $2\,\mathrm{km/h}$. \\
Independent random wake & A delay sampled from $U(15,45)$ simulation seconds, unaffected by the Driving-Agent heartbeat. \\
Coalescing & A five-second cooldown merges repeated events by category. A PA may also finish without issuing a request; its update still wakes the Driving Agent once at that boundary. \\
\bottomrule
\end{tabular}
\end{table}

Judge checks, tool receipts, request expiry or non-completion, and DA
heartbeats do not generate PA wakes.

\subsubsection{Typed deferred-request triggers}

Table~\ref{tab:trigger-catalog} lists the supported trigger families. The
runtime filters candidates using the remaining route, scheduled environmental
events, and free-flow time available for the Judge window. Numeric parameters
are bounded by the current state; free-text predicates are not supported.

\begingroup
\normalsize
\setlength{\LTleft}{0pt}
\setlength{\LTright}{0pt}
\begin{longtable}{@{}>{\raggedright\arraybackslash}p{31mm}@{\hspace{3mm}}>{\raggedright\arraybackslash}p{\dimexpr\columnwidth-34mm\relax}@{}}
\caption{Declarative Judge-trigger catalog. Only currently feasible,
parameter-bounded members of this catalog are shown to the PA.}
\label{tab:trigger-catalog}\\
\toprule
Family & Available predicates \\
\midrule
\endfirsthead
\toprule
Family (continued) & Available predicates \\
\midrule
\endhead
\bottomrule
\endfoot
Time & A bounded delay after request creation. \\
Route and map & Approaching, entering, or exiting the next intersection; distance to destination below a bounded threshold. \\
Vehicle state & Vehicle stopped, vehicle resumed moving, or a speed threshold held for a bounded duration. \\
Environment & A scheduled weather transition to an exposed condition, or a scheduled transition to darkness. \\
\end{longtable}
\endgroup

Selected predicates are checked at the 0.1-second physics resolution.

\subsubsection{Request lifecycle and Judge windows}

The request contract records core and secondary criteria, a one-shot or
ongoing kind, expected response time, validity duration, and an optional
trigger. Judge checks occur at offsets of 0.1, 1, and 3 seconds after immediate
creation or trigger activation, with at most three checks per phase. Intermediate
checks retain criterion status; the final verdict assigns one grade.
Evidence includes the contract, DA replies, tool receipts, before/after state,
and the physical trace.

\begingroup
\normalsize
\setlength{\LTleft}{0pt}
\setlength{\LTright}{0pt}
\begin{longtable}{@{}>{\raggedright\arraybackslash}p{31mm}@{\hspace{3mm}}>{\raggedright\arraybackslash}p{\dimexpr\columnwidth-34mm\relax}@{}}
\caption{Request lifecycle. Terminal states are reported as coverage rather
than silently folded into the request score.}
\label{tab:request-lifecycle}\\
\toprule
State & Meaning \\
\midrule
\endfirsthead
\toprule
State (continued) & Meaning \\
\midrule
\endhead
\bottomrule
\endfoot
Created / waiting & The frozen request exists; a deferred request is waiting for a feasible selected predicate. \\
Activated / checking & The predicate occurred and the Judge gathers the bounded time-window evidence. \\
Completed & Evidence establishes fulfillment within the request validity period; one-shot requests may close early once all core criteria are established. \\
Uncompleted / unverified & The check budget ends with unmet criteria or insufficient evidence, respectively. \\
Superseded / episode ended & A later PA request replaces the request, or the physical episode ends before its pending lifecycle is resolved. \\
NA / infrastructure failure & An invalid request or insufficient evidence is unscored (NA); operational failures are tracked separately. \\
\end{longtable}
\endgroup

Ongoing requests remain pending until the final observation window.
Verdict validation checks criterion status, phase boundaries, timestamps, and
trace consistency; malformed verdicts remain evaluation failures.

\begin{algorithm}[H]
\normalsize
\caption{Passenger Request and Verification Loop at Each World Boundary}
\label{alg:pa-judge-loop}
\begin{minipage}[t]{\columnwidth}
\vspace{0pt}
\begin{algorithmic}[1]
\Require World state $s$, events $e$, active request sheet $\mathcal{S}$, recorded execution history
\State $p\gets\emptyset$ \Comment{only passenger updates are returned to DA}
\If{$\operatorname{PAIsDue}(s,e)$}
    \State $\Omega\gets\operatorname{AvailableTriggers}(s,\operatorname{RemainingRoute},\operatorname{TimeBudget})$
    \State $d\gets\operatorname{PA}(\operatorname{PassengerView}(s),\mathcal{S},\Omega)$
    \State Validate structured request updates and freeze their acceptance criteria
    \State $(\mathcal{S},p)\gets\operatorname{ApplyAcceptedUpdates}(\mathcal{S},d)$
\EndIf
\For{each pending request $r\in\mathcal{S}$}
    \State Update $r$'s verification schedule from its trigger, deadline, and episode status
    \If{$\operatorname{CheckDue}(r)$}
        \State $x\gets\operatorname{CollectExecutionEvidence}(r)$
        \State $v\gets\operatorname{VerifyRequest}(r,x,\operatorname{PriorChecks}(r))$
        \State Record a valid verdict, or retain an explicit unscored status
        \State Close $r$ if complete or at its final check; otherwise retain it for a later check
    \EndIf
\EndFor
\State \Return $p$ \Comment{no driver wake is generated by verification}
\end{algorithmic}
\end{minipage}
\end{algorithm}

\Needspace{8\baselineskip}
\subsection{Evaluation Metric Definitions}
\label{app:metric-definitions}

Delay increases are summed per vehicle, so time saved by one vehicle does not offset delays imposed on another. Trip-completion delay is computed only for vehicles that arrive in the evaluated run. Untriggered or unverified requests and failed Judge calls are reported separately, not assigned grades.

\begingroup
\normalsize
\setlength{\LTleft}{0pt}
\setlength{\LTright}{0pt}
\begin{longtable}{@{}>{\raggedright\arraybackslash}p{31mm}@{\hspace{3mm}}>{\raggedright\arraybackslash}p{29mm}@{\hspace{3mm}}>{\raggedright\arraybackslash}p{\dimexpr\columnwidth-66mm\relax}@{}}
\caption{VehicleArena metrics. Metrics are reported separately rather than
combined into one scalar.}
\label{tab:benchmark-metrics}\\
\toprule
Metric & Unit and denominator & Definition and companion report \\
\midrule
\endfirsthead
\toprule
Metric (continued) & Unit and denominator & Definition and companion report \\
\midrule
\endhead
\bottomrule
\endfoot
Trip completion $A$ & Percentage of tasks & Ego arrival by $T_{\mathrm{limit}}$ without collision or route failure. Peer arrivals are reported separately, not required for ego success. \\
Driving quality $D$ & Score in $[0,100]$ per evaluated vehicle & Starts at 100 with deductions for driving violations; an at-fault collision or red-light violation sets the score to zero. \\
Passenger fulfillment $R$ & Score in $[0,100]$ over valid scored requests & A--F Judge grades mapped to 100, 80, 60, 40, 20, and 0. Request-level and scenario-level means use explicit denominators. \\
Cabin compliance $C$ & Score in $[0,100]$ over checked fields & Field-weighted accuracy pooled across ego checkpoints, including sentinel checks. Rules requiring unavailable equipment are excluded. \\
Externality $E$ & Counts and vehicle-seconds per valid pair & Changes in NPC collisions and non-arrivals, plus positive increases in queueing time and trip-completion time relative to SUMO. \\
Efficiency $K$ & Input and output tokens per task & Total model-token use, reported separately for input and output. \\
\end{longtable}
\endgroup

\subsection{Passenger Evaluation and Reporting}
\label{app:metrics}

\subsubsection{Aggregation and unscored cases}

For each aggregate, report its valid observation count and the numbers of
unavailable, unscored, or invalid cases. An empty applicable set is reported
as unavailable, not as a zero score.

The request-level mean pools all valid scored requests. The scenario-level
mean first averages valid request scores within each scenario, then averages
over scenarios with at least one scored request. Thus, the latter gives each
applicable scenario equal weight rather than weighting it by request count.

\subsubsection{Passenger-request grades}

Table~\ref{tab:request-grades} expands the six grades in
Table~\ref{tab:benchmark-metrics}. Judge assesses only the frozen passenger
requirements, not unrelated driving quality or traffic impact. Timing is based
on evidence of fulfillment, not the time at which Judge returns its verdict.

\begin{table}[!htb]
\caption{Passenger-request grading rubric. NA is an unscored status, not a seventh grade.}
\label{tab:request-grades}
\centering
\normalsize
\begin{tabularx}{\columnwidth}{>{\raggedright\arraybackslash}p{10mm}rX}
\toprule
Grade & Score & Evidence-based interpretation \\
\midrule
A & 100 & All requirements are resolved correctly within the expected response time. \\
B & 80 & All requirements are resolved, but after the expected response time and before expiry. \\
C & 60 & All core requirements are resolved, with minor secondary omissions or missing secondary evidence. \\
D & 40 & Useful partial fulfillment of core requirements, or failure to maintain a sustained requirement. \\
E & 20 & Relevant physical action is observed, but it produces no useful requested outcome. \\
F & 0 & No relevant action, an unrelated action, or an outcome contrary to the request. \\
NA & --- & An invalid request or evidence insufficient for a fair judgment; excluded from score averages. \\
\bottomrule
\end{tabularx}
\end{table}

A requirement is resolved by observed fulfillment or a clear, truthful refusal
supported by evidence that the vehicle lacks the requested capability. Such a
refusal does not establish physical completion. Promises, Todo edits, and
queued commands alone do not establish fulfillment; an already satisfied state
can count. Sustained requirements are checked across the observation window.

\subsubsection{Traffic Interaction Set}

The traffic-interaction set contains all required non-ego trip vehicles,
including fixed peers in \textsc{MultiLLM}. Evaluated and SUMO reference runs
are paired by task and vehicle ID. Queueing delay, collision, and non-arrival
measures use this full set; trip-completion delay is computed only for
vehicles arriving in both runs. Figure~\ref{fig:surrounding-impact} uses the
same scope, with the measures defined in Section~\ref{app:surrounding-impact}.

\subsection{Surrounding-Vehicle Impact Measures}
\label{app:surrounding-impact}

Figure~\ref{fig:surrounding-impact} uses all required non-ego trip vehicles,
including fixed peers in Multi-Agent tasks. For each model and task group,
let $\mathcal{T}$ contain the 80 Basic or 32 Multi-Agent tasks, and let $V_t$
be the surrounding trip vehicles in task $t$. We pair each evaluated run $M$
with its SUMO reference $S$ by task and vehicle ID. Basic uses an all-SUMO
reference; Multi-Agent uses a SUMO ego vehicle with fixed peers.

Let $a_{tv}^{X}$ be arrival time, $q_{tv}^{X}$ queue-wait time, and
$r_{tv}^{X}\in\{0,1\}$ the arrival indicator for vehicle $v$ in run
$X\in\{M,S\}$. Let $c_{tv}^{X}\in\{0,1\}$ indicate whether that vehicle was
involved in at least one collision in run $X$. Define
$V_t^{\cap}=\{v\in V_t:r_{tv}^{M}=r_{tv}^{S}=1\}$,
$N_{\mathrm{trip}}=\sum_{t\in\mathcal{T}}|V_t|$, and
$[x]_+=\max(x,0)$. The four baseline-relative measures are
\begin{equation}
\begin{aligned}
D_{\mathrm{arr}} &= \frac{1}{|\mathcal{T}|}\sum_{t\in\mathcal{T}}\sum_{v\in V_t^{\cap}}
  [a_{tv}^{M}-a_{tv}^{S}]_+, &
D_{\mathrm{queue}} &= \frac{1}{|\mathcal{T}|}\sum_{t\in\mathcal{T}}\sum_{v\in V_t}
  [q_{tv}^{M}-q_{tv}^{S}]_+,\\
\Delta C &= \frac{100}{N_{\mathrm{trip}}}\sum_{t\in\mathcal{T}}\sum_{v\in V_t}
  (c_{tv}^{M}-c_{tv}^{S}), &
\Delta N &= \frac{100}{N_{\mathrm{trip}}}\sum_{t\in\mathcal{T}}\sum_{v\in V_t}
  (r_{tv}^{S}-r_{tv}^{M}).
\end{aligned}
\end{equation}

The delays $D_{\mathrm{arr}}$ and $D_{\mathrm{queue}}$ have units of
vehicle-seconds per task. $\Delta C$ and $\Delta N$ are treatment-minus-SUMO
changes in collision and non-arrival rates, in percentage points. A vehicle
counts once in $\Delta C$ even if it has multiple collision events. The radar
axes plot
$(100D_{\mathrm{arr}}/45,\;100D_{\mathrm{queue}}/60,\;
100[\Delta C]_+/1.25,\;\Delta N)$.
The delay divisors are 45 and 60 vehicle-seconds per task; the collision
divisor is 1.25 percentage points. Because the radar cannot show a negative
radius, a collision-rate decrease is displayed at zero on that axis; the
signed difference remains in the underlying data. Non-arrival is shown as
the baseline difference in percentage points without further scaling. For
descriptive comparisons, we average these four plotted coordinates and then
average the Basic and Multi-Agent values with equal weight; this is a visual
summary, not an additional evaluation metric.

\subsection{Completion-Weighted Outcome Index}
\label{app:completion-weighted-index}

For each model and task group, $A$, $D$, $R$, and $C$ denote the Arr. (\%),
Drive, Req., and Cabin metrics in Table~\ref{tab:main-results}, respectively.
We use their unrounded values and write $A$ as a percentage (e.g., $A=65\%$).
The plotted index is
\begin{equation}
I = A\,\frac{D+R+C}{3}.
\label{eq:completion-weighted-index}
\end{equation}

\section{Supplementary Analyses}
\label{app:supplementary-analyses}

\subsection{Model-Level Skill, Tool, and Token Counts}
\label{app:model-preference-counts}

Table~\ref{tab:model-preference-counts} gives the counts behind
Figure~\ref{fig:model-skill-tool} and the ego DA token totals for the same
models. Each row pools 80 Basic and 32 Multi-Agent tasks. Skill entries count
tasks with a successful load of the named skill; tool entries count non-\texttt{finish}
calls to each tool family. Day/night, Weather, Driving, and Safety denote
\texttt{daynight\_transition}, \texttt{weather\_transition},
\texttt{driving\_control}, and \texttt{safety\_refusal}. The tool columns
Nav., API, Front, A/C, Todo, and Rear
denote \texttt{navigation}, \texttt{get\_module\_api}, \texttt{frontRadar},
\texttt{airConditioner}, \texttt{todo\_manage}, and \texttt{rearRadar},
respectively. Other collects the remaining tool families. Token entries are
total ego DA input and output tokens across the 112 tasks, reported in
millions (M; $1\,\mathrm{M}=10^6$ tokens) and rounded to two decimal places.

\begin{table}[!htb]
\caption{Per-model skill loads, tool calls, and ego DA token counts.}
\label{tab:model-preference-counts}
\centering
\normalsize
\setlength{\tabcolsep}{3pt}
\renewcommand{\arraystretch}{1.06}
\begin{tabular*}{\textwidth}{@{\extracolsep{\fill}}lrrrr@{}}
\toprule
\multicolumn{5}{l}{Successful skill loads (tasks)} \\
\midrule
Model & Day/night & Weather & Driving & Safety \\
\midrule
DeepSeek-V4.1-Flash & 51 & 41 & 62 & 5 \\
MIMO-V2.6-Pro & 37 & 35 & 7 & 0 \\
Kimi-K3 & 14 & 23 & 2 & 0 \\
GLM-5.3-Flash & 30 & 30 & 4 & 0 \\
Qwen3.8-Max & 46 & 48 & 0 & 1 \\
Qwen3.8-27B & 26 & 19 & 1 & 1 \\
GPT-5.6-Sol & 57 & 50 & 105 & 9 \\
Qwen3-VL-8B & 47 & 51 & 23 & 2 \\
Qwen3-VL-32B & 30 & 26 & 1 & 0 \\
\bottomrule
\end{tabular*}

\vspace{0.8em}
\begin{tabular*}{\textwidth}{@{\extracolsep{\fill}}lrrrrrrr@{}}
\toprule
\multicolumn{8}{l}{Tool-family calls} \\
\midrule
Model & Nav. & API & Front & A/C & Todo & Rear & Other \\
\midrule
DeepSeek-V4.1-Flash & 2,717 & 1,467 & 1,669 & 442 & 1,085 & 1,093 & 5,681 \\
MIMO-V2.6-Pro & 1,658 & 953 & 687 & 467 & 1,329 & 563 & 3,549 \\
Kimi-K3 & 1,681 & 1,046 & 645 & 439 & 787 & 513 & 2,747 \\
GLM-5.3-Flash & 1,289 & 897 & 588 & 458 & 624 & 562 & 2,848 \\
Qwen3.8-Max & 1,772 & 1,000 & 559 & 461 & 817 & 624 & 3,225 \\
Qwen3.8-27B & 1,336 & 859 & 588 & 426 & 685 & 450 & 2,248 \\
GPT-5.6-Sol & 4,495 & 1,286 & 1,627 & 544 & 1,052 & 1,176 & 5,653 \\
Qwen3-VL-8B & 5,243 & 525 & 209 & 2,898 & 356 & 443 & 6,065 \\
Qwen3-VL-32B & 7,877 & 993 & 1,674 & 1,425 & 575 & 1,067 & 4,235 \\
\bottomrule
\end{tabular*}

\vspace{0.8em}
\begin{tabular*}{\textwidth}{@{\extracolsep{\fill}}lrrr@{}}
\toprule
\multicolumn{4}{l}{Ego DA tokens (M)} \\
\midrule
Model & Input & Output & Total \\
\midrule
DeepSeek-V4.1-Flash & 331.64 & 43.89 & 375.53 \\
MIMO-V2.6-Pro & 162.80 & 15.63 & 178.43 \\
Kimi-K3 & 117.53 & 7.69 & 125.22 \\
GLM-5.3-Flash & 121.58 & 8.73 & 130.31 \\
Qwen3.8-Max & 141.16 & 12.25 & 153.41 \\
Qwen3.8-27B & 108.83 & 8.86 & 117.68 \\
GPT-5.6-Sol & 265.82 & 8.94 & 274.76 \\
Qwen3-VL-8B & 234.35 & 5.89 & 240.24 \\
Qwen3-VL-32B & 301.43 & 6.17 & 307.60 \\
\bottomrule
\end{tabular*}
\end{table}

\subsection{Token Cost Analysis}
\label{app:token-cost-analysis}

\begin{figure}[!htbp]
\centering
\includegraphics[width=0.83\textwidth]{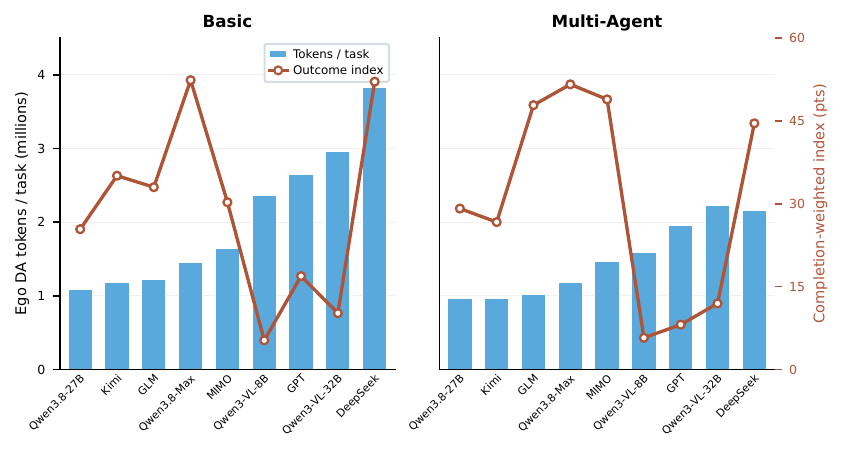}
\caption{Inference cost and task outcome by model (Basic: left; Multi-Agent:
right). Bars show mean ego DA tokens per task (left axis, millions); the line
shows the completion-weighted index (right axis, points).}
\label{fig:model-token-cost}
\end{figure}

For the cost comparison in Figure~\ref{fig:model-token-cost}, we count ego DA
input and output tokens per task. Passenger agents, judges, and fixed peer
models are excluded. Both panels use the same model order, set by Basic token
use, and the same scales.

Token totals add a different view of effort. Runs with similar wake and tool
counts can process different amounts of text during each wake, so interaction
counts need not predict inference use. We therefore compare tokens with the
same completion-weighted outcome to see whether additional processing is
accompanied by better task results.

On Multi-Agent, GLM-5.3-Flash substantially outperforms Kimi-K3 and
Qwen3.8-27B at nearly the same token budget, while Qwen3.8-Max reaches the
highest outcome with only moderately more tokens. At a higher, roughly
two-million-token budget, DeepSeek-V4.1-Flash outperforms GPT-5.6-Sol. Thus,
token volume alone does not explain task success. We use token count as a
proxy for inference cost without accounting for differences in provider
pricing.

Figure~\ref{fig:model-token-cost} shows that similar token budgets can yield
very different outcomes. On Basic, Qwen3.8-Max achieves the strongest outcome
among models using roughly one to two million ego DA tokens per task; it also
matches DeepSeek-V4.1-Flash while using less than half as many tokens.

\section{Passenger Requests in the PA Ablation}
\label{app:pa-low-speed-examples}

Table~\ref{tab:pa-low-speed-examples} expands four representative ego
trajectories into six individual passenger requests, paired with the road
and environment context at the time of each request. Times are simulation
seconds. Requests are translated or abridged, with unrelated device clauses
omitted. The numerical speeds are passenger preferences, not road speed
limits; distance thresholds specify when the passenger wants an action to
occur, not the measured remaining distance when the request is issued.

\begin{table}[!htb]
\caption{Scene context and low-speed passenger requests in the PA ablation.}
\label{tab:pa-low-speed-examples}
\centering
\apptablestyle
\begin{tabularx}{\textwidth}{@{}>{\raggedright\arraybackslash}p{0.46\textwidth}>{\raggedright\arraybackslash}X@{}}
\toprule
Scene at request time & Passenger request \\
\midrule
Urban lead-vehicle following in Beijing, in cloudy weather. At 5.0 s,
a hard-braking event prompts a request for more separation from the car ahead.
& After that hard brake, slow to about 25 km/h so we keep more distance
from the car ahead. \\
\addlinespace[4pt]
The same following episode, after rain begins. At 20.4 s, the passenger
updates rain-related device settings and plans a quiet final approach.
& Once we are within about 130 m of the destination, slow to about
20 km/h for the final approach and turn the music off. \\
\addlinespace[6pt]
Platoon-following traffic in Shanghai. At 14.1 s, fog appears as afternoon
changes to dusk, reducing visibility.
& It is foggy and we are approaching the destination; slow to about
20 km/h. \\
\addlinespace[4pt]
The same foggy episode at dusk. At 19.1 s, a later request specifies a
further slowdown for the final approach and preparation for arrival.
& When navigation shows less than 70 m to the destination, slow to about
10 km/h and pause the music to prepare for arrival. \\
\addlinespace[6pt]
An urban crosswalk route in Shanghai at dusk. Rain begins at 15.1 s;
the passenger requests wipers, fog lights, and a cautious final approach.
& Once we are within 50 m of the destination, slow to below 15 km/h,
then stop smoothly after arrival. \\
\addlinespace[6pt]
An urban road with oncoming traffic in Sydney, in the afternoon. At
10.0 s, a passenger who had been reading asks for quieter audio and
prepares to pack up.
& Within 100 m of the destination, approach gently at under 10 km/h
and turn off the passenger reading light so I can pack up. \\
\bottomrule
\end{tabularx}
\end{table}

\section{Driving-Rule Scoring Details}
\label{app:driving-rules}

Driving quality starts at 100. An at-fault collision or red-light violation
sets the score to zero; otherwise, the score is
$\max(0,100-\sum_j p_j)$, where $p_j$ is a recorded deduction in
Table~\ref{tab:driving-deductions}. These are event-level penalties, not
weights of a normalized average. Weather-dependent cabin compliance is
scored separately rather than deducted again here.

\begin{table}[!htb]
\caption{Driving-score deduction schedule. Repeated detections are grouped
into events or episodes before scoring.}
\label{tab:driving-deductions}
\centering
\normalsize
\begin{tabularx}{\columnwidth}{X>{\raggedleft\arraybackslash}p{28mm}}
\toprule
Violation & Deduction \\
\midrule
At-fault collision or red-light violation & Score becomes 0 \\
Sustained speeding, at least 5 seconds & 20 \\
Short speeding episode, 1 to less than 5 seconds & 5 \\
Unjustified stopping & 5 each \\
Unjustified delay in starting after a green light & 5 \\
Lane change without signaling & 10 \\
Turn without signaling & 10 \\
Failure to cancel a turn signal & 2 \\
Unsafe lane change & 10 \\
Late response to a lead-vehicle, junction, or pedestrian risk & 10 each \\
Vehicle near miss & 15 \\
Junction-conflict near miss & 15 \\
Pedestrian near miss & 20 \\
Blocking an intersection & 10 \\
Harsh acceleration & 3 \\
Unnecessary harsh braking & 3 \\
High longitudinal jerk & 1 \\
Unnecessary horn use & 2 \\
High-beam misuse & 2 \\
Entering a closed road & 10 \\
\bottomrule
\end{tabularx}
\end{table}

Deduction types have a three-second cooldown. Continuous unjustified stops
and intersection blocking incur an initial penalty after three seconds and
repeat at five-second intervals while the condition persists.

Driving deductions use observed physical behavior. Pedestrian near misses
are identified from vehicle--pedestrian body clearance and vehicle speed,
rather than predicted arrival timing alone. Late responses and near misses
within one continuous pedestrian encounter incur only the most severe
deduction, preventing repeated penalties for the same encounter.

\section{Vehicle Modules and Configuration}
\label{app:configuration-rules}

The task configuration resolves a separate immutable capability set for every
vehicle before the first simulation step. Table~\ref{tab:vehicle-configuration}
lists the supported fields. A scenario may provide defaults for all vehicles
and overrides for an individual vehicle. Validation occurs before the episode
starts, so an unknown module or invalid chassis value cannot become a
mid-episode tool failure.

\begin{table}[!htb]
\caption{Per-vehicle configuration fields.}
\label{tab:vehicle-configuration}
\centering
\normalsize
\begin{tabularx}{\columnwidth}{>{\raggedright\arraybackslash}p{31mm}X}
\toprule
Field & Effect \\
\midrule
Equipment profile & Selects a named set of cabin and sensor modules. \\
enable\_modules / disable\_modules & Adds or removes named optional modules for one vehicle. Required world and driving modules cannot be removed. \\
Chassis profile & Selects vehicle length, width, acceleration, braking, and lane-change duration. \\
Chassis overrides & Replaces individual positive chassis values for a task-specific vehicle variant. \\
Trusted extension & At run initialization, registers additional modules, equipment profiles, chassis profiles, scenario layers, or rule directories. Scenario data cannot import code. \\
\bottomrule
\end{tabularx}
\end{table}

The same resolved capability set controls module instantiation, the initial
tool catalog, the available skill catalog, and physical limits. If DA tries to
inspect or invoke an uninstalled module, the interface returns an explicit
capability\_not\_available result. This makes an unsupported request
distinguishable from a failed attempt to use an installed device.

\subsection{Built-in module catalog}

Table~\ref{tab:module-catalog} lists the 40 modules registered by the current
release. These are the modules that a profile may expose; an individual
vehicle receives only the subset selected by its capability set. The five
world and driving modules are required by the runtime. The others are optional
and can be included, excluded, or supplemented by a trusted extension.

\begin{table}[!htb]
\caption{Built-in VehicleArena modules. Names are the module identifiers used
by the capability resolver and tool-discovery interface.}
\label{tab:module-catalog}
\centering
\normalsize
\begin{tabularx}{\textwidth}{>{\raggedright\arraybackslash}p{31mm}X}
\toprule
Group & Registered modules \\
\midrule
World and driving & navigation, map, weather, dayNight, speedLimit \\
Sensors & frontRadar, rearRadar, lidar \\
Climate and comfort & airConditioner, seat, sunshade, readingLight \\
Display, media, and communication & bluetooth, broadcast, centerInformationDisplay, conversation, HUD, instrumentPanel, music, overheadScreen, radio, video, horn \\
Lighting and safety & fogLight, hazardLight, highBeamHeadlight, lowBeamHeadlight, positionLight, tailLight, turnSignal, wiper \\
Body and physical controls & door, footPedal, frontTrunk, fuelPort, rearviewMirror, steeringWheel, sunroof, trunk, window \\
\bottomrule
\end{tabularx}
\end{table}

\subsection{Declarative rule format and evaluation}
\label{app:cabin-rule-evaluation}

Cabin requirements are authored as validated YAML rules rather than embedded
in evaluator code. Each rule has an identifier and one domain-specific trigger,
an optional guard, expected actions, and optional tolerances. The supported
domains are weather transitions, day--night transitions, map events, and user
intent. Tolerances can name accepted alternatives, fields to ignore, a required
numeric trend, or a numerical ceiling. Negative checks specify a state that
must not occur. Unknown fields or invalid action syntax are rejected when the
rule files are loaded.

In the reported experiments, automatic Cabin expectations are derived from
weather, day--night, and map transitions. Although the schema supports user
intent, PA-generated requests are evaluated separately by the Passenger Judge
and do not generate user-intent Cabin checkpoints in either PA-on or PA-off.
Disabling the ego PA/Judge removes ego request generation and judgment; fixed
peers retain both roles.

At a relevant transition, the rule engine derives the expected action set from
the current and previous world snapshots. For scoring, those actions run on an
isolated reference VehicleWorld. The cabin evaluator compares the
reference state change with the actual state change of the evaluated vehicle,
checking the target fields affected by the reference actions with the declared
tolerances. Rules requiring uninstalled modules are excluded. An already
satisfied expectation counts as one fulfilled check. A newly violated negative
invariant, or a checkpoint for which all reference candidates fail to execute,
contributes one failed check.
The reported Cabin score is
\(100\,\sum\texttt{correct\_fields}/\sum\texttt{total\_fields}\), pooled
across the ego checkpoints in the reported task set. The number of checked
fields can vary between PA-on and PA-off because reference-action state changes
depend on the current device state. Free-form DA text and an API call without
its required state change do not count as compliance.

\section{Supported Cities}
\label{app:cities}
Table~\ref{tab:city-coverage} lists map areas represented in the configured
scenario catalog by region and city. Evaluation tasks use a subset of these areas.

{\normalsize
\begin{longtable}{>{\raggedright\arraybackslash}p{25mm}>{\raggedright\arraybackslash}p{31mm}>{\raggedright\arraybackslash}p{71mm}}
\caption{City-level map inventory. Each local area corresponds to one
lane-level map; the table does not imply that every map appears in the
112-task evaluation set.}
\label{tab:city-coverage} \\
\toprule
Region & City & Local area(s) \\
\midrule
\endfirsthead
\multicolumn{3}{l}{\normalsize\textit{Table~\ref{tab:city-coverage} continued.}} \\
\toprule
Region & City & Local area(s) \\
\midrule
\endhead
\midrule
\multicolumn{3}{r}{\normalsize\textit{Continued on next page.}} \\
\endfoot
\bottomrule
\endlastfoot
Mainland China (57) & Beijing & Guomao; Sanlitun; Tiananmen; Wangjing; Wudaokou; Xidan; Yizhuang; Zhongguancun \\
& Changchun & Chaoyang \\
& Changsha & Furong; Wuyi \\
& Changzhou & Tianning \\
& Chengdu & Chunxi; Gaoxin; Jinjiang \\
& Chongqing & Jiefangbei \\
& Dalian & Zhongshan \\
& Dongguan & Nancheng \\
& Foshan & Chancheng \\
& Fuzhou & Wuyi \\
& Guangzhou & Panyu; Tianhe \\
& Guiyang & Guanshanhu \\
& Haikou & Longhua \\
& Hangzhou & Binjiang; Xihu \\
& Harbin & Zhongyang \\
& Hefei & Shushan; Zhengwu \\
& Hohhot & Saihan \\
& Jiaxing & Nanhu \\
& Jinan & Quancheng \\
& Kunming & Cuihu \\
& Lanzhou & Chengguan \\
& Lhasa & Chengguan \\
& Nanchang & Honggutan \\
& Nanjing & Hexi; Xinjiekou \\
& Nantong & Chongchuan \\
& Ningbo & Tianyi \\
& Qingdao & Shinan; Wusi \\
& Shanghai & Hongkou; Jing'anbei; Lujiazui; Pudong Zhangjiang \\
& Shenyang & Heping; Zhongjie \\
& Shenzhen & Futian; Nanshan \\
& Suzhou & Guanqian \\
& Taiyuan & Yingze \\
& Tianjin & Heping \\
& Wuhan & Hankou \\
& Wuxi & Taihu \\
& Xiamen & Huli \\
& Xi'an & Zhonglou \\
\midrule
East and Southeast Asia (10) & Bangkok & Silom \\
& Hanoi & Hoankiem \\
& Hong Kong & Central \\
& Jakarta & Central \\
& Kuala Lumpur & Bukit \\
& Osaka & Namba \\
& Seoul & Gangnam \\
& Singapore & Orchard \\
& Taipei & Xinyi \\
& Tokyo & Shinjuku \\
\midrule
Europe (13) & Amsterdam & Centrum \\
& Berlin & Mitte \\
& Helsinki & Keskusta \\
& Istanbul & Beyoglu \\
& London & West End \\
& Madrid & Centro \\
& Moscow & Tverskaya \\
& Paris & Champs \\
& Prague & Stare Mesto \\
& Rome & Centro \\
& Stockholm & Norrmalm \\
& Vienna & Innere \\
& Warsaw & Srodmiescie \\
\midrule
North America (5) & Chicago & Loop \\
& Los Angeles & Downtown \\
& New York & Manhattan Midtown \\
& San Francisco & SoMa \\
& Toronto & Downtown \\
\midrule
Oceania (2) & Melbourne & CBD \\
& Sydney & CBD \\
\midrule
Middle East and Africa (2) & Cairo & Downtown \\
& Dubai & Downtown \\
\end{longtable}
}

\Needspace{5\baselineskip}
\section{Model Specifications}
\label{app:model-signatures}

Table~\ref{tab:model-signatures} summarizes the nine models evaluated in our
experiments.
Peer DA, PA, and Judge all use Qwen3.8-27B.

\begin{table}[!htb]
\caption{Specifications of the large language models evaluated in our experiments.}
\label{tab:model-signatures}
\centering
\normalsize
\setlength{\tabcolsep}{3pt}
\renewcommand{\arraystretch}{1.16}
\begin{tabular*}{\columnwidth}{@{\extracolsep{\fill}}lccrcll@{}}
\toprule
Models & \#Para & Launch date & \shortstack{Context\\(tokens)} & Scaling & Corporation & License \\
\midrule
DeepSeek-V4.1-Flash & 748B & 2026-09-10 & 1,048,576 & Effort & DeepSeek & MIT \\
MiMo-v2.6-Pro & 1.02T & 2026-09-22 & 1,048,576 & Budget & Xiaomi & MIT \\
Kimi-K3 & 2.8T & 2026-07-17 & 1,048,576 & Effort & Moonshot AI & Kimi K3 \\
GLM-5.3-Flash & 320B & 2026-09 & 1,048,576 & Effort & Z.ai & MIT \\
Qwen3.8-Max & 2.4T & 2026-08-02 & 1,000,000 & Effort & Alibaba & Proprietary \\
Qwen3.8-27B & 27B & 2026-08-14 & 262,144\textsuperscript{a} & Effort & Alibaba & Apache 2.0 \\
GPT-5.6-sol & --- & 2026-07-09 & 1,050,000 & Effort & OpenAI & Proprietary \\
Qwen3-VL-8B & 8.8B & 2025-10-15 & 262,144\textsuperscript{a} & Budget & Alibaba & Apache 2.0 \\
Qwen3-VL-32B & 33.4B & 2025-10-21 & 262,144\textsuperscript{a} & Budget & Alibaba & Apache 2.0 \\
\bottomrule
\end{tabular*}
\par\vspace{3pt}
\begin{minipage}{\columnwidth}
\normalsize
\textsuperscript{a}Native context, extensible to 1M tokens.
Context is model capacity; the experimental output limit is 32,768 tokens
(Table~\ref{tab:inference-settings}). Effort/Budget denotes
reasoning-effort/generation-length control. Qwen3-VL uses Instruct
checkpoints; counts include the vision encoder. Licenses apply to released
weights where available; Qwen3.8-Max is the hosted model.
\end{minipage}
\end{table}


\section{Prompts}
\label{app:prompts}

This section presents the prompt templates used by the Driving Agent,
Personal Agent, and Judge in VehicleArena.

\lstdefinestyle{prompttemplate}{
  basicstyle=\rmfamily\normalsize\selectfont,
  frame=none,
  breaklines=true,
  breakatwhitespace=false,
  breakautoindent=false,
  breakindent=0pt,
  columns=fullflexible,
  keepspaces=true,
  showstringspaces=false,
  upquote=true,
  literate={"}{{\fontencoding{T1}\selectfont\textquotedbl}}1 {-}{{-}}1,
  aboveskip=0pt,
  belowskip=0pt
}

\newtcblisting{appendixcontent}[2][]{
  enhanced jigsaw,
  breakable,
  lines before break=4,
  listing only,
  listing options={style=prompttemplate},
  colback=white,
  colbacktitle=black,
  colframe=gray!60,
  coltitle=white,
  fonttitle=\bfseries\normalsize,
  title={#2},
  title after break={#2},
  arc=2mm,
  boxrule=0.4pt,
  titlerule=0pt,
  left=8pt,
  right=8pt,
  top=6pt,
  bottom=6pt,
  toptitle=3pt,
  bottomtitle=3pt,
  before skip=8pt,
  after skip=10pt,
  #1
}

\Needspace{6\baselineskip}
\begin{appendixcontent}{Driving Agent Prompt}
SYSTEM
You directly control one vehicle in a continuous road-world simulation. Use tools for driving and cabin operations. The world advances in 0.1-second steps. A target-speed command persists; accepted motion commands are not completed maneuvers. Use CameraVisual for visible traffic and signal state. Use installed radar and LidarBEV only when available. The assigned destination remains your long-term Todo goal. Route planning supplies guidance, not automatic route following. Choose legal junction maneuvers and signal lane changes yourself. Use get_module_api and load_tools for installed capabilities. Use set_heartbeat_interval or schedule_next_wake to observe again. Call finish when this wake is complete.
<installed-module catalog and chassis limits>
<available skill catalog>
<multi-vehicle context: vehicle_id>
<driving-role addendum: driver_prompt, if configured>

USER AT EACH WAKE
CurrentWake: <time, ego state, active control, wake policy,
              Todo board, passenger updates, new events>
CameraVisual: <aligned cockpit image>
LidarBEV: <image only if LiDAR is installed>
LoadedCapabilities: <startup or changed callable schemas>
PreviousWake: <at most one retained exchange if budget permits>
\end{appendixcontent}

\Needspace{6\baselineskip}
\begin{appendixcontent}{Personal Agent Prompt}
SYSTEM
You are the human passenger riding in this vehicle. React only to passenger-observable cabin, weather, motion, traffic, comfort, and safety state. Do not control equipment directly or request a change to the assigned destination. Call finish when no request is warranted. Otherwise submit one natural passenger utterance with explicit, observable core outcomes and genuinely optional secondary outcomes. Use send_immediate_request for outcomes beginning now. Use send_triggered_request only for a two-stage request with immediate and triggered outcomes and an offered judge trigger. Give request_kind, expected_response_s, and valid_for_s. If an active sheet is replaced, restate every still-desired outcome; cancel it only when none remains wanted.

USER AT PA WAKE
{
  "type": "personal_agent_wake",
  "persona": <passenger persona>,
  "todo": {"long_term_goal": <passenger goal>},
  "observation": <passenger-visible state, active request sheet,
                  request design, available judge triggers>
}
TOOLS
send_immediate_request | send_triggered_request |
cancel_passenger_request | finish
\end{appendixcontent}

\Needspace{6\baselineskip}
\begin{appendixcontent}{Judge Prompt}
SYSTEM
You are an independent VehicleArena passenger-request judge.
<passenger grade rubric and physical-outcome contract>
Judge only the frozen requested outcomes using observed evidence. An accepted command, Todo edit, or promise alone is not proof of physical fulfillment. Previously committed control may explain observed motion. Classify each core and secondary criterion as met, unsupported_refused, unmet, or unverified. Use unsupported_refused only for an explicit passenger-facing refusal backed by authoritative unavailability evidence. For triggered criteria, measure time from trigger activation; for immediate criteria, measure time from request creation. At an intermediate window, submit criterion statuses without an A--F grade. At the final window, submit one evidence-backed grade or explicit NA status and fulfillment times for resolved phases. Judge ongoing behavior within the observed window.

USER AT JUDGE CHECK
<frozen request and criteria; phase/timing metadata; DA replies;
 tool receipts; before/after device state; physical trace;
 prior checks; intermediate or final window flag>

TOOL AT INTERMEDIATE WINDOW: submit_passenger_check
TOOL AT FINAL WINDOW: submit_passenger_judgement
\end{appendixcontent}

\section{Skills}
\label{app:skills}

Skills are optional Markdown guides discovered by DA through the catalog
embedded in its system instruction. Calling \path{load_skill(skill_name)}
loads the named guide into retained context. The shipped catalog contains the four
guides in Table~\ref{tab:skill-catalog}. Guidance naming equipment absent
from the current vehicle is filtered when loaded. The weather guide expands a
shared set of weather-safety rules at load time.

\begin{table}[!htb]
\caption{Operational skill catalog in the implementation.}
\label{tab:skill-catalog}
\centering
\normalsize
\begin{tabularx}{\columnwidth}{>{\raggedright\arraybackslash}p{41mm}X}
\toprule
Skill & Use \\
\midrule
\mbox{\textbf{\texttt{driving\_control}}} & Motion commands, perception, routing, and command receipts. \\
\mbox{\textbf{\texttt{weather\_transition}}} & Equipment changes for weather and visibility. \\
\mbox{\textbf{\texttt{daynight\_transition}}} & Lights, displays, and mirrors around darkness. \\
\mbox{\textbf{\texttt{safety\_refusal}}} & Unsafe-request refusal and safe alternatives. \\
\bottomrule
\end{tabularx}
\end{table}

The following listings show the complete guide bodies before vehicle-specific
equipment filtering, with the shared weather rules expanded. YAML frontmatter
is omitted; typography is normalized, and the intersection-name example is
translated into English.

\Needspace{8\baselineskip}
\subsection{\texorpdfstring{\texttt{driving\_control}}{driving\_control}}
\label{app:skill-driving-control}

\begin{appendixcontent}{Skill: \texttt{driving\_control}}
# Driving Control Tool Reference

This skill explains what the driving tools do. It does not choose a driving personality, desired speed, risk tolerance, following distance, signal compliance, or maneuver for the driver.

## Continuous execution

- The physical world advances in 0.1-second steps.
- A successful control call means the command was accepted for the next world commit. It does not mean the maneuver completed instantly.
- Target speed persists until replaced.
- Commands submitted in one wake are committed as one batch. Commands that control the same motion slot do not stack: the later command wins and the earlier command receipt reports `superseded`.
- Acceleration and braking are bounded.
- A lane change is a continuous lateral trajectory lasting several seconds.
- Intersections use explicit source-lane -> connector -> destination-lane paths.
- Vehicle rectangles and pedestrian circles are checked with swept collision geometry throughout each step.

## Perception

Every vehicle wake includes `CameraVisual`, captured from the synchronized Web3D cockpit. It is the optical source for visible lanes, road users, signal lights and vehicle lamps, and is affected by weather, daylight and lighting. An installed optional LiDAR adds `LidarBEV`: the established ego-centred rule-rendered top-down geometry image. It deliberately omits traffic-light state, lamp effects, weather and day/night appearance. Vehicles without LiDAR receive no substitute image. Installed front/rear millimetre-wave radar tools provide numeric anonymous vehicle tracks. The visual heartbeat defaults to one second; the driver may set it between 0.1 and 30 simulation seconds. Genuine onboard radar warnings may interrupt it; evaluator-only world-risk labels never wake the driver.

## Route planning

`navigation_route_plan(destination)` resolves a destination to a suggested lane route for display. It returns only a compact success receipt; inspect the route geometry with `navigation_minimap`. Planning neither moves the vehicle nor selects a SUMO intersection connector.

The mini-map is heading-up with the ego vehicle near the lower centre. `scope="route"` provides a wider forward driving window and `scope="local"` provides a closer junction/lane window. Long routes leave the image boundary; they are not compressed into a whole-trip overview.

Destination names may identify:

- an intersection, such as `"Road A & Road B"`;
- a POI or landmark on a road.

Arrival is committed by the physical world after contact/collision checks, not merely because a tool was called.

For an intended left/right/U-turn, call `navigation_select_maneuver("left" | "straight" | "right" | "u_turn")` before entering the junction. An explicit straight selection is also supported. One successful call selects exactly one legal source-lane -> connector -> destination-lane movement. A direction unavailable from the current lane is rejected; change lanes explicitly and wait for completion before trying again. The selection remains active through that junction and is not cancelled by an intermediate wake. After exiting that junction, the selection is consumed; a left turn does not make the vehicle keep choosing left at later junctions.

Without an explicit selection, the vehicle continues through the current lane's unique legal straight connector, if one exists. This is lane continuation, not automatic following of the highlighted navigation route. If there is no unique straight connector, the system does not brake or choose a turn for you; an unresolved route endpoint can terminate the trip as a failure. Observe the road and decide in time. Background NPC routes remain fully SUMO-controlled. If the highlighted route terminates on the current lane, continue to its endpoint without selecting another junction maneuver.

Traffic signals are movement-specific: a straight green arrow does not permit a left turn whose arrow is red. Observe the signal for your intended movement.

## Longitudinal commands

`navigation_set_speed(...)` submits a persistent target speed plus optional acceleration and ordinary-deceleration limits. SUMO applies the command under the installed chassis bounds. Speed zero requests an ordinary physical stop.

`navigation_emergency_stop(reason)` requests the installed chassis emergency braking envelope. It still takes physical time.

There are no `follow`, `overtake` or `yield` modes. Express those decisions through explicit speed and lane-change commands, then observe their physical result.

## Lane changes and U-turns

`navigation_change_lane("left" | "right")` starts a physical lane-change trajectory when the requested lane exists. Target-lane bodies remain present, and a collision can occur during the maneuver.

The physical actuator does not add an indicator automatically. Before starting a lane change, call the preloaded `turnSignal__switch` tool with the matching direction. Switch it off after the maneuver finishes. Other agents can perceive the emitted indicator subject to their optical range and conditions.

`navigation_u_turn()` is the dedicated form of selecting a U-turn connector from the current lane. It does not change lanes, rotate, or relocate the vehicle automatically.

## Observable road rules and consequences

Signals, limits, lane markings, crosswalks, and right-of-way are observable rules. The simulator records violations and physical consequences but leaves the behavior decision to the driver.

A collision disables the involved vehicle. The wreck remains a physical obstacle on its actual lane or connector; adjacent lanes can remain usable.

## Wake events

`CurrentWake.new_events` contains non-visual facts such as onboard radar warnings, command receipts, passenger requests and terminal state. Visual traffic facts are never delivered as semantic wake events. `active_control` contains only ego commands that actually committed and remain relevant.

After `simulation_ended`, newly submitted physical commands are discarded.
\end{appendixcontent}

\Needspace{8\baselineskip}
\subsection{\texorpdfstring{\texttt{weather\_transition}}{weather\_transition}}
\label{app:skill-weather-transition}

\begin{appendixcontent}{Skill: \texttt{weather\_transition}}
# Weather Transition Rules

**Only act when you see a `[Weather]` event.** Do not preemptively adjust settings -- the vehicle's defaults are correct for clear weather.

## Required external equipment

Apply the entering-weather settings and any applicable cleanup below. Only operate installed equipment; an already satisfied state needs no repeat action. These instructions use the same rules as scoring and NPCs.

### Entering `foggy`

- `lowBeamHeadlight.switch('on')`
- `highBeamHeadlight.switch(False)`
- `fogLight.carcontrol_fogLight_switch(True, 'front')`
- `fogLight.carcontrol_fogLight_switch(True, 'rear')`
- `positionLight.carcontrol_positionLight_switch(True)`

### Entering `hail`

- `window.carcontrol_window_switch(['all'], False)`
- `sunroof.carcontrol_sunroof_switch('close')`

### Entering `heavy_rain`

- `wiper.carcontrol_wiperBlade_switch(True, 'front')`
- `window.carcontrol_window_switch(['all'], False)`
- `sunroof.carcontrol_sunroof_switch('close')`

### Entering `heavy_snow`

- `wiper.carcontrol_wiperBlade_switch(True, 'front')`
- `window.carcontrol_window_switch(['all'], False)`
- `sunroof.carcontrol_sunroof_switch('close')`
- `lowBeamHeadlight.switch('on')`
- `highBeamHeadlight.switch(False)`
- `fogLight.carcontrol_fogLight_switch(True, 'front')`
- `fogLight.carcontrol_fogLight_switch(True, 'rear')`
- `positionLight.carcontrol_positionLight_switch(True)`

### Entering `rainy`

- `wiper.carcontrol_wiperBlade_switch(True, 'front')`
- `window.carcontrol_window_switch(['all'], False)`
- `sunroof.carcontrol_sunroof_switch('close')`

### Entering `snowy`

- `wiper.carcontrol_wiperBlade_switch(True, 'front')`
- `window.carcontrol_window_switch(['all'], False)`
- `sunroof.carcontrol_sunroof_switch('close')`
- `lowBeamHeadlight.switch('on')`

### Leaving `heavy_rain`, `heavy_snow`, `rainy`, `snowy` for a condition outside that set

- `wiper.carcontrol_wiperBlade_switch(False, 'front')`

### Leaving `foggy`, `heavy_snow` for a condition outside that set

- `fogLight.carcontrol_fogLight_switch(False, 'front')`
- `fogLight.carcontrol_fogLight_switch(False, 'rear')`
- `positionLight.carcontrol_positionLight_switch(False)`

During dusk/night/dawn, keep positionLight ON despite weather cleanup; the day/night lighting requirement takes precedence. Weather cleanup does not turn lowBeamHeadlight off. Weather cleanup does not reopen window or sunroof. Sunny/cloudy adds no entering-weather action; apply only relevant cleanup.

## Cold weather (-> snowy/heavy_snow/hail)
- Turn ON steering wheel heater
- Turn ON rearview mirror heating
- Turn ON seat heater

## Low visibility (-> foggy/heavy_rain/snowy/heavy_snow/hail)
- Enable HUD and reduce brightness

## Leaving fog
- Turn OFF defrost and auto-defog

## Leaving cold weather
- Turn OFF steering wheel heater, mirror heating, seat heater

## Leaving low visibility
- Restore HUD brightness
\end{appendixcontent}

\Needspace{8\baselineskip}
\subsection{\texorpdfstring{\texttt{daynight\_transition}}{daynight\_transition}}
\label{app:skill-daynight-transition}

\begin{appendixcontent}{Skill: \texttt{daynight\_transition}}
# Day/Night Transition Rules

**Only act when you see a `[DayNight]` event.** Do not preemptively adjust settings based on the current time of day -- the vehicle's defaults are already correct for daytime.

## Entering dark period (morning/noon/afternoon -> dusk/night/dawn)

**All four actions are required:**
1. `lowBeamHeadlight.switch('on')` -- headlights on
2. `positionLight.carcontrol_positionLight_switch(True)` -- position lights on
3. `centerInformationDisplay.brightness_decrease(degree='large')` -- dim display
4. `rearviewMirror.mode_autoAdjust(True)` -- anti-glare

## Leaving dark period (dusk/night/dawn -> morning/noon/afternoon)
- Turn OFF low beam headlights: `lowBeamHeadlight.switch('off')`
- Turn OFF position lights: `positionLight.carcontrol_positionLight_switch(False)`
- Increase center display brightness: `centerInformationDisplay.brightness_increase(degree='large')`
- Disable rearview mirror auto-adjust: `rearviewMirror.mode_autoAdjust(False)`

## First tick in dark period
- Same as entering dark period (treat as initial setup)
\end{appendixcontent}

\Needspace{8\baselineskip}
\subsection{\texorpdfstring{\texttt{safety\_refusal}}{safety\_refusal}}
\label{app:skill-safety-refusal}

\begin{appendixcontent}{Skill: \texttt{safety\_refusal}}
# Safety Refusal Rules

## When to refuse
Refuse and broadcast a safety warning when the passenger requests:
- Opening doors while driving
- Opening trunk while driving
- Playing video while driving (distraction)
- Any action that compromises vehicle safety

## How to refuse
- Do NOT execute the unsafe action
- Call `broadcast.broadcast_safety_refusal(True, reason)` with a clear reason
- Respond politely to the passenger explaining why the request cannot be fulfilled

## HARD safety constraints (enforced by ConstraintEngine)
These are blocked at the system level regardless:
- Door open while speed > 0
- Trunk open while speed > 0
- Video playback while driving

## Safe alternatives
- If passenger wants to open door: suggest stopping first
- If passenger wants video: offer audio-only alternatives (music, radio)
\end{appendixcontent}

\section{Tool Inventory}
\label{app:tools}

\subsection{Driving Agent (DA) Tools}
\label{app:da-tools}

DA tool schemas come from installed modules and the harness. The ten core
calls are listed in Table~\ref{tab:core-tools}; discovery and harness calls
appear in Table~\ref{tab:da-harness-tools}. Other \path{module__method} calls
are inspected and loaded on demand, with loaded schemas retained across
wakes subject to the context budget.

\newcommand{\toolentry}[2]{%
  \mbox{\normalsize\textbf{\texttt{\detokenize{#1}}}} & #2\\%
}
\begingroup
\normalsize
\setlength{\LTleft}{0pt}
\setlength{\LTright}{0pt}
\begin{longtable}{@{}>{\raggedright\arraybackslash}p{86mm}@{\hspace{3mm}}>{\raggedright\arraybackslash}p{\dimexpr\columnwidth-89mm\relax}@{}}
\caption{Core DA tools preloaded when their modules are installed.}
\label{tab:core-tools}\\
\toprule
Tool & Function \\
\midrule
\endfirsthead
\toprule
Tool (continued) & Function \\
\midrule
\endhead
\toolentry{navigation__navigation_route_plan}{Display a suggested route.}
\toolentry{navigation__navigation_minimap}{Show a route or local map without replanning.}
\toolentry{navigation__navigation_set_speed}{Set target speed and optional acceleration limits.}
\toolentry{navigation__navigation_emergency_stop}{Request emergency braking.}
\toolentry{navigation__navigation_change_lane}{Request one adjacent lane.}
\toolentry{navigation__navigation_select_maneuver}{Select a legal junction connector.}
\toolentry{frontRadar__scan}{Read the front radar.}
\toolentry{rearRadar__scan}{Read the rear radar.}
\toolentry{speedLimit__speed_limit_get}{Read the current speed limit.}
\toolentry{turnSignal__switch}{Set a persistent turn signal.}
\bottomrule
\end{longtable}
\endgroup

DA discovers APIs for the installed subset of the modules listed in
Table~\ref{tab:module-catalog}; absent equipment is omitted. The discovery
catalog also includes \path{road_perception}, a virtual engine API. Its text tools omit
camera-visible objects and signals; optional \path{lidar} supplies an image.

\begingroup
\normalsize
\setlength{\LTleft}{0pt}
\setlength{\LTright}{0pt}
\begin{longtable}{@{}>{\raggedright\arraybackslash}p{70mm}@{\hspace{3mm}}>{\raggedright\arraybackslash}p{\dimexpr\columnwidth-73mm\relax}@{}}
\caption{DA discovery, harness, and additional tool functions.}
\label{tab:da-harness-tools}\\
\toprule
Tool & Function \\
\midrule
\endfirsthead
\toprule
Tool (continued) & Function \\
\midrule
\endhead
\toolentry{get_module_api}{List a module's callable methods.}
\toolentry{load_tools}{Load selected module APIs.}
\toolentry{load_skill}{Load procedural guidance.}
\toolentry{todo_manage}{Maintain tasks across wakes.}
\toolentry{memory_search}{Search passenger, event, and action history.}
\toolentry{set_heartbeat_interval}{Set periodic observation interval ($0.1$--$30$ s).}
\toolentry{schedule_next_wake}{Schedule a replaceable one-shot observation.}
\toolentry{get_device_state}{Read cabin equipment state.}
\toolentry{finish}{End the current DA wake.}
\toolentry{navigation__navigation_u_turn}{Request an optional U-turn.}
\bottomrule
\end{longtable}
\endgroup

Cabin calls report state changes; motion calls report acceptance and need
later receipts or observation to confirm completion. Invalid calls return
repairable errors.

\subsection{Personal Agent (PA) Tools}
\label{app:pa-tools}

PA's four non-driving calls (Table~\ref{tab:pa-tools}) manage the passenger
request sheet shared with DA and Judge.

\begin{table}[!htb]
\caption{PA request-management tools.}
\label{tab:pa-tools}
\centering
\normalsize
\begin{tabularx}{\columnwidth}{@{}>{\raggedright\arraybackslash}p{65mm}@{\hspace{3mm}}>{\raggedright\arraybackslash}X@{}}
\toprule
Tool & Function \\
\midrule
\toolentry{send_immediate_request}{Create or replace an immediate request with outcomes and timing.}
\toolentry{send_triggered_request}{Create or replace a two-phase request with a typed trigger.}
\toolentry{cancel_passenger_request}{Cancel the active request when no pending outcome is wanted.}
\toolentry{finish}{End the wake; keep the active request unchanged.}
\bottomrule
\end{tabularx}
\end{table}

\subsection{Judge Tools}
\label{app:judge-tools}

Judge's submissions (Table~\ref{tab:judge-tools}) are validated against the
frozen request and timeline; neither changes vehicle state or wakes DA.

\begin{table}[!htb]
\caption{Judge submission tools.}
\label{tab:judge-tools}
\centering
\normalsize
\begin{tabularx}{\columnwidth}{@{}>{\raggedright\arraybackslash}p{65mm}@{\hspace{3mm}}>{\raggedright\arraybackslash}X@{}}
\toprule
Tool & Function \\
\midrule
\toolentry{submit_passenger_check}{Record intermediate criterion statuses and evidence; no grade.}
\toolentry{submit_passenger_judgement}{Submit a final A--F/NA grade, statuses, and fulfillment times.}
\bottomrule
\end{tabularx}
\end{table}

\end{document}